\documentclass[lettersize,journal]{IEEEtran}

\usepackage[thmmarks]{ntheorem} 
\usepackage{amsmath,amsfonts}
\usepackage{algorithmic}
\usepackage{algorithm}
\usepackage{array}
\usepackage{textcomp}
\usepackage{todonotes}
\usepackage{stfloats}
\usepackage{url}
\usepackage{verbatim}
\usepackage{graphicx}
\usepackage{epsfig}
\usepackage{cite}
\usepackage{pifont}
\usepackage{amsmath}
\usepackage{booktabs}
\usepackage{subfigure} 
\usepackage{makecell}
\usepackage{subfloat}
\usepackage{bm}
\usepackage{xcolor}
\usepackage{tabularx} 
\usepackage{ragged2e} 
\usepackage{longtable}
\usepackage{supertabular}

\usepackage{array}
\newcolumntype{C}[1]{>{\centering\arraybackslash}p{#1}} 
\newcolumntype{L}[1]{>{\raggedright\arraybackslash}p{#1}}

\usepackage[titles,subfigure]{tocloft}
\makeatletter
\renewcommand{\maketag@@@}[1]{\hbox{\m@th\normalsize\normalfont#1}}%
\makeatother

\usepackage{nomencl} 

\begin{document}
	
	\title{
		\textcolor{black}{Dominant Transient Stability of the Co-located PLL-Based Grid-Following Renewable Plant and Synchronous Condenser Systems}}
	\author{Bingfang Li,~\IEEEmembership{Member,~IEEE},
			Songhao Yang,~\IEEEmembership{Senior Member,~IEEE},
			Qinglan Wang,\\ Xu Zhang,~\IEEEmembership{Graduate Student Member,~IEEE}, Huan Xie, Chuan Qin, Zhiguo Hao,~\IEEEmembership{Senior Member,~IEEE}
	\thanks{This work was supported by the Key Research and Development Program of Shaanxi (2025PT-ZCK-01). The authors acknowledge J. Guo from the Shaanxi Key Laboratory of Smart Grid for providing access to the instruments that supported this work.}
	\thanks{B. Li, S. Yang, Q. Wang, X. Zhang, and Z. Hao are with Xi'an Jiaotong University, Xi'an 710049, China (e-mail: libingfang@stu.xjtu.edu.cn; songhaoyang@xjtu.edu.cn; zhghao@xjtu.edu.cn).}
	\thanks{H. Xie and C. Qin are with State Grid Jibei Electric Power Co., Beijing 100000, China.}}
	\maketitle
\begin{abstract}
	\textcolor{black}{Deploying synchronous condensers (SynCons) near grid-following renewable energy sources (GFLRs) is an effective and increasingly adopted strategy for grid support. However, the potential transient instability risks in such configurations remain an open research question. This study investigates the mechanism of dominant synchronization instability source transition upon SynCon integration and proposes a straightforward approach to enhance system stability by leveraging their interactive characteristics.
		Firstly, a dual-timescale decoupling model is established, partitioning the system into a fast subsystem representing phase-locked loop (PLL) dynamics and a slow subsystem characterizing SynCon rotor dynamics. The study then examines the influence of SynCons on the transient stability of nearby PLLs and their own inherent stability. The study shows that SynCon's voltage-source characteristics and its time-scale separation from PLL dynamics can significantly enhance the PLL's stability boundary and mitigate non-coherent coupling effects among multiple GFLRs. However, the dominant instability source shifts from the fast-time-scale PLL to the slow-time-scale SynCon after SynCon integration. Crucially, this paper demonstrates that the damping effect of PLL control can also be transferred from the fast to the slow time scale, allowing well-tuned PLL damping to suppress SynCon rotor acceleration. Consequently, by utilizing SynCon's inherent support capability and a simple PLL damping loop, the transient stability of the co-located system can be significantly enhanced. These conclusions are validated using a converter controller-based Hardware-in-the-Loop (CHIL) platform.}
\end{abstract}

\section*{Nomenclature}
\begin{IEEEdescription}[\IEEEsetlabelwidth{$R_{ci}, R_s, R_g$}]
	\item[\textcolor{black}{$\delta$}] \textcolor{black}{Rotor angle of SynCon relative to the grid.}
	\item[\textcolor{black}{$\theta_i$}] \textcolor{black}{Output angle of the $i$-th PLL relative to the grid.}
	\item[\textcolor{black}{$\omega_s, \omega_{ci}, \omega_g$}] \textcolor{black}{Angular frequency of the SynCon, the $i$-th converter, the grid.}
	\item[\textcolor{black}{$\Delta\omega, \varpi_{ci}$}] \textcolor{black}{Angular frequency of SynCon and the $i$-th converter relative to the grid.}
	\item[\textcolor{black}{$E_s \angle \delta$}] \textcolor{black}{Electric potential of the SynCon.}
	\item[\textcolor{black}{$U_{ci} \angle \phi_{ci}$}] \textcolor{black}{Voltage phasor of the $i$-th converter.}
	\item[\textcolor{black}{$U_g \angle 0^{\circ}$}] \textcolor{black}{Voltage phasor of the grid bus.}
	\item[\textcolor{black}{$I_{ci}\angle\varphi_{ci}$}] \textcolor{black}{Current phasor of the $i$-th converter.}
	\item[\textcolor{black}{$I_{s}\angle\varphi_{ci}$}] \textcolor{black}{Current phasor of the SynCon.}
	\item[\textcolor{black}{$L_{ci}, L_s, L_g$}] \textcolor{black}{GFLR, SynCon, and grid-side branch inductance.}
	\item[\textcolor{black}{$R_{ci}, R_s, R_g$}] \textcolor{black}{GFLR, SynCon, and grid-side branch resistance.}
	\item[\textcolor{black}{$Y_{ci}, Y_s, Y_g$}] \textcolor{black}{GFLR, SynCon, and grid-side branch admittance.}
	\item[\textcolor{black}{$\alpha$}] \textcolor{black}{Coupling coefficient of GFLR and SynCon, defined as $Y_s/(Y_s+Y_g)$.}
	\item[\textcolor{black}{$d_i\!-\!q_i$}] \textcolor{black}{Rotating reference frame of the $i$-th converter.}
	\item[\textcolor{black}{$d_s\!-\!q_s$}] \textcolor{black}{Rotating reference frame of the SynCon.}
	\item[\textcolor{black}{$x\!-\!y$}] \textcolor{black}{Rotating reference frame of the grid.}
	\item[\textcolor{black}{$k_{Pi}, k_{Ii}$}] \textcolor{black}{Proportional, integral gain of the $i$-th PLL.}
	\item[\textcolor{black}{$i_{ci}^{drf}, i_{ci}^{qrf}$}] \textcolor{black}{Active, reactive current references of the $i$-th converter.}
	\item[\textcolor{black}{$i_{ci}^d, i_{ci}^q$}] \textcolor{black}{Active and reactive current of the $i$-th converter.}
	\item[\textcolor{black}{$P_E$}] \textcolor{black}{Real electromagnetic power of SynCon.}
	\item[\textcolor{black}{$T_s$}] \textcolor{black}{Inertia time constant of the SynCon.}
	\item[\textcolor{black}{$D_{s}$}] \textcolor{black}{Damping coefficient of the SynCon.}
	\item[\textcolor{black}{$P_{c}$}] \textcolor{black}{Equivalent mechanical power of the SynCon.}
	\item[\textcolor{black}{$P_{es}$}] \textcolor{black}{Equivalent electromagnetic power of SynCon.}
	\item[\textcolor{black}{$u_{ci}^{q}$}] \textcolor{black}{The $q_i$-axis component of $U_{ci} \angle \phi_{ci}$.}
	\item[\textcolor{black}{$i_{cij}^{d_i}$}] \textcolor{black}{The sum of the $d_i$-axis components of the currents from all other converters.}
	\item[\textcolor{black}{$T_{ci}$}] \textcolor{black}{Equivalent time constant of the $i$-th PLL.}
	\item[\textcolor{black}{$P_{Mci}, P_{Eci}$}] \textcolor{black}{Equivalent mechanical power, equivalent electromagnetic  power of the $i$-th PLL.}
	\item[\textcolor{black}{$D_{ci}$}] \textcolor{black}{Equivalent damping coefficient of the $i$-th PLL with respect to the synchronous grid.}
	\item[\textcolor{black}{$D_{cij}$}] \textcolor{black}{Equivalent damping term of the $i$-th PLL with respect to other PLLs.}
	\item[\textcolor{black}{$t, \tau$}] \textcolor{black}{Slow, fast time scale time variable.}
	\item[\textcolor{black}{$\varepsilon$}] \textcolor{black}{Small parameter for time scale separation.}
	\item[\textcolor{black}{$\overline{\delta}$}] \textcolor{black}{Frozen value of $\delta$ on fast time scale.}
	\item[\textcolor{black}{$i_{cij}^{d_i*}$}] \textcolor{black}{The $d_i$-axis component of the currents from other converters under coherent condition.}
	\item[\textcolor{black}{$U_E$}] \textcolor{black}{An equivalent voltage after SynCon integration.}
	\item[\textcolor{black}{$\theta_\delta$}] \textcolor{black}{Arctan term of the SEP angle $\theta_{si}^*$.}
	\item[\textcolor{black}{$\theta_{bi}$}] \textcolor{black}{The stability index for the $i$-th PLL.}
	\item[\textcolor{black}{$\theta_i^{'}$}] \textcolor{black}{Transformed phase angle, defined as $\theta_i - \theta_\delta$.}
	\item[\textcolor{black}{$P_{Mci}^*$}] \textcolor{black}{Equivalent mechanical power of the $i$-th converter under multi-converter coherent operation.}
	\item[\textcolor{black}{$\Delta P_{Mci}$}] \textcolor{black}{Defined as $P_{Mci}-P_{Mci}^*$.}
	\item[\textcolor{black}{$D_{ci}^*$}] \textcolor{black}{Damping coefficient of the $i$-th converter under multi-converter coherent operation.}
	\item[\textcolor{black}{$\Delta D_{ci}$}] \textcolor{black}{Defined as $D_{ci}-D_{ci}^*$.}
	\item[\textcolor{black}{$u_i$}] \textcolor{black}{Non-coherent disturbance term.}
	\item[\textcolor{black}{$u_i^{\rm{max}}, u_i^{\rm{min}}$}] \textcolor{black}{Maximum, minimum non-coherent disturbance term of the $i$-th converter.}
	\item[\textcolor{black}{$I_{cj}^{\rm{max}}$}] \textcolor{black}{Maximum allowable current for the $j$-th GFLC.}
	\item[\textcolor{black}{$V_{ci}, \dot{V}_{ci}$}] \textcolor{black}{Transient energy function and its time derivative of the $i$-th converter.}
	\item[\textcolor{black}{$\theta_{si}^*$}] \textcolor{black}{SEP angle of the $i$-th converter.}
	\item[\textcolor{black}{$\eta_{ci}$}] \textcolor{black}{Power factor angle of the $i$-th converter current, defined as $\theta_i - \varphi_{ci}$.}
	\item[\textcolor{black}{$i_c^{d_s}$}] \textcolor{black}{Sum of the projections of each GFLC current onto the $d_s$-axis.}
	\item[\textcolor{black}{$\Delta\delta$}] \textcolor{black}{Change in the $\delta$ during a small time interval $\Delta t$.}
	\item[\textcolor{black}{$\Delta\theta_i$}] \textcolor{black}{Change in the angle $\theta_i$ of the $i$-th converter during a small time interval $\Delta t$.}
	\item[\textcolor{black}{$t_0, t_1$}] \textcolor{black}{Instants post-PLL synchronization and during \(\delta\) acceleration in the first swing after \(t_0\).}
	\item[\textcolor{black}{$t_f$}] \textcolor{black}{Estimated fault duration time.}
	\item[\textcolor{black}{$K_d$}] \textcolor{black}{Additional Damping loop coefficient of PLL.}
\end{IEEEdescription}

	\section{Introduction}
	\IEEEPARstart{R}{e}newable energy sources, such as wind and solar power, are undergoing rapid growth. Most of these resources are integrated into the grid through grid-following converters (GFLCs) for grid connection\cite{1}. However, GFLCs possess inherent limitations, including restricted grid-support capabilities and susceptibility to tripping under disturbances\cite{2}. 
	\textcolor{black}{Since large-scale renewable generation is often located remotely from load centers, these plants typically serve as sending-end systems. The transmission of bulk power over  long electrical distances inherently weakens the AC grid connection. This fragility exacerbates stability risks, contributing to wide-band oscillations, violations of frequency and voltage limits, and, in extreme cases, large-scale blackouts\cite{4,5}.}
	
	Against this backdrop, there is a growing need for renewable generation facilities to develop grid-supporting capabilities. \textcolor{black}{Two main approaches have emerged as promising solutions. The first approach is introducing grid-forming (GFM) control, which is predominantly based on the concept of the virtual synchronous generators (VSGs)\cite{6}.  VSGs emulate the inertial response and damping characteristics of synchronous generators (SGs), thus providing essential grid-support services\cite{8}. Nevertheless, due to the inherently low overcurrent capability of power electronic devices, GFM units are susceptible to entering current-limiting mode during transients\cite{9}. This not only constrains their ability to support the grid but also introduces more complex stability challenges. }
	
	\textcolor{black}{The second approach involves combining grid-following renewable energy sources (GFLRs) with synchronous condensers (SynCons). This solution has been discussed in \cite{11}, and many studies have shown that such configuration can mitigate transient overvoltage and frequency violations, while reducing the risk of wide-band oscillations \textcolor{black}{in renewable energy systems}\cite{4,14,15,16}. Moreover, SynCons offer a significant advantage over power-electronic-based devices in their superior overcurrent withstand capability, enabling them to maintain a voltage source characteristic even during severe system disturbances. }

	\textcolor{black}{Globally, this configuration is being actively planned and adopted in several countries. Energinet, the transmission system operator (TSO) of Denmark, has emphasized the value of SynCons in improving system stability and delivering ancillary services \cite{14}. Nordic TSOs likewise recognize that adding rotating inertia from SynCons can help address the low-inertia challenge \cite{14}. The State Grid Corporation of China (SGCC) has explicitly advocated for the deployment of SynCons at renewable energy sites to bolster short-circuit capacity support in its national grid development plans. \cite{18}. In practice, distributed SynCons have been deployed and put into operation at several renewable energy facilities across China \cite{22,23}. Operational data indicate that their integration has increased renewable output from approximately 30\% to more than 70\% of installed capacity \cite{23}. Similar applications have also been reported in the United States, where SynCons have been used to enhance voltage, frequency, and reactive power support \cite{25}. These experiences collectively suggest that the ``GFLR+SynCon" configuration is a promising solution for strengthening current and future power system architectures. }
	
	\textcolor{black}{In conventional understanding, SynCons, as reactive compensation devices without prime movers, appear to be at little risk of transient acceleration issues like those in SGs. They are only likely to experience decelerating instability when located at load centers\cite{40}.}  However, as noted in \cite{26} and \cite{27}, the GFLR output can be equivalently regarded as the mechanical power input to the SynCon, thereby providing the impetus for angle acceleration. In addition, refs.\cite{26} and \cite{28} examine the influence of different fault types and low-voltage ride-through (LVRT) strategies on SynCon instability modes and transient energy accumulation. Their findings indicate that, under certain conditions, SynCons may exhibit an unusual instability pattern, decelerating during short circuits and subsequently undergoing acceleration instability after fault clearance. \textcolor{black}{These investigations primarily address the stability of supporting devices \textcolor{black}{within GFLR grid connected systems}, leaving two critical issues that merit further exploration:}
	
	\textcolor{black}{First, phase-locked loops (PLLs) in GFLC are widely recognized for their inherent fragility, particularly during voltage disturbances \cite{12,29,31,32}. Considering the recent occurrences of instability in synchronous condensers, what is the dominant transient instability issue: the PLL or the SynCon?}

		\textcolor{black}{Second, how should SynCon instability be effectively addressed? The most straightforward option is to trip SynCons or a portion of GFLRs. However, this compromises the structural integrity of the network and, furthermore, tripping these units may trigger a secondary large disturbance. Enhancing SynCon stability through excitation control \cite{36} faces limitations, particularly in adapting to time-varying disturbances from nearby GFLRs. 
		A potentially more fundamental solution is to target GFLR output, the root cause of SynCon instability. This concept has been investigated in \cite{28}, where system frequency and output power are measured in real time and fed back to GFLR plants for output adjustment. Nevertheless, significant practical challenges remain. These include the intricate coordination of multiple GFLR units with diverse operating states, and the imperative of maintaining energy balance and DC link voltage stability following output adjustments. }
		
	\textcolor{black}{To answer these questions, this study investigates the impact mechanism of SynCon integration on the transient stability of the original GFLC system, and the transfer mechanism of the system's dominant transient  instability source.  Furthermore, it proposes a practical method for stability enhancement. The main contributions are as follows:}
	\begin{enumerate}
		\item This study develops a dual time-scale transient stability analysis model, separating the system into a fast subsystem (representing PLL dynamics) and a slow subsystem (representing SynCon rotor dynamics). This framework provides a clear basis for analyzing the impact mechanism of SynCon integration on system stability.
		\item \textcolor{black}{This study reveals the mechanism by which the system's dominant instability source changes after SynCon integration. The SynCon's voltage-source characteristics and the decoupling of its rotor dynamics from PLL dynamics significantly enhance the transient stability of GFLC-based systems and mitigate non-coherent inter-converter coupling effects.  However, the primary instability source transitions from the GFLR’s PLL to the SynCon.}
		\item \textcolor{black}{This study reveals that well-tuned PLL damping control is capable of damping the rotor acceleration of the SynCon. Building on this finding, it offers a simple and practical engineering approach for enhancing dual-time-scale transient stability.}
	\end{enumerate}
	
	\section{{System Modeling}}
	\subsection{System Overview}
\begin{figure}[!t]
		\centering
		\includegraphics[width=3.4in]{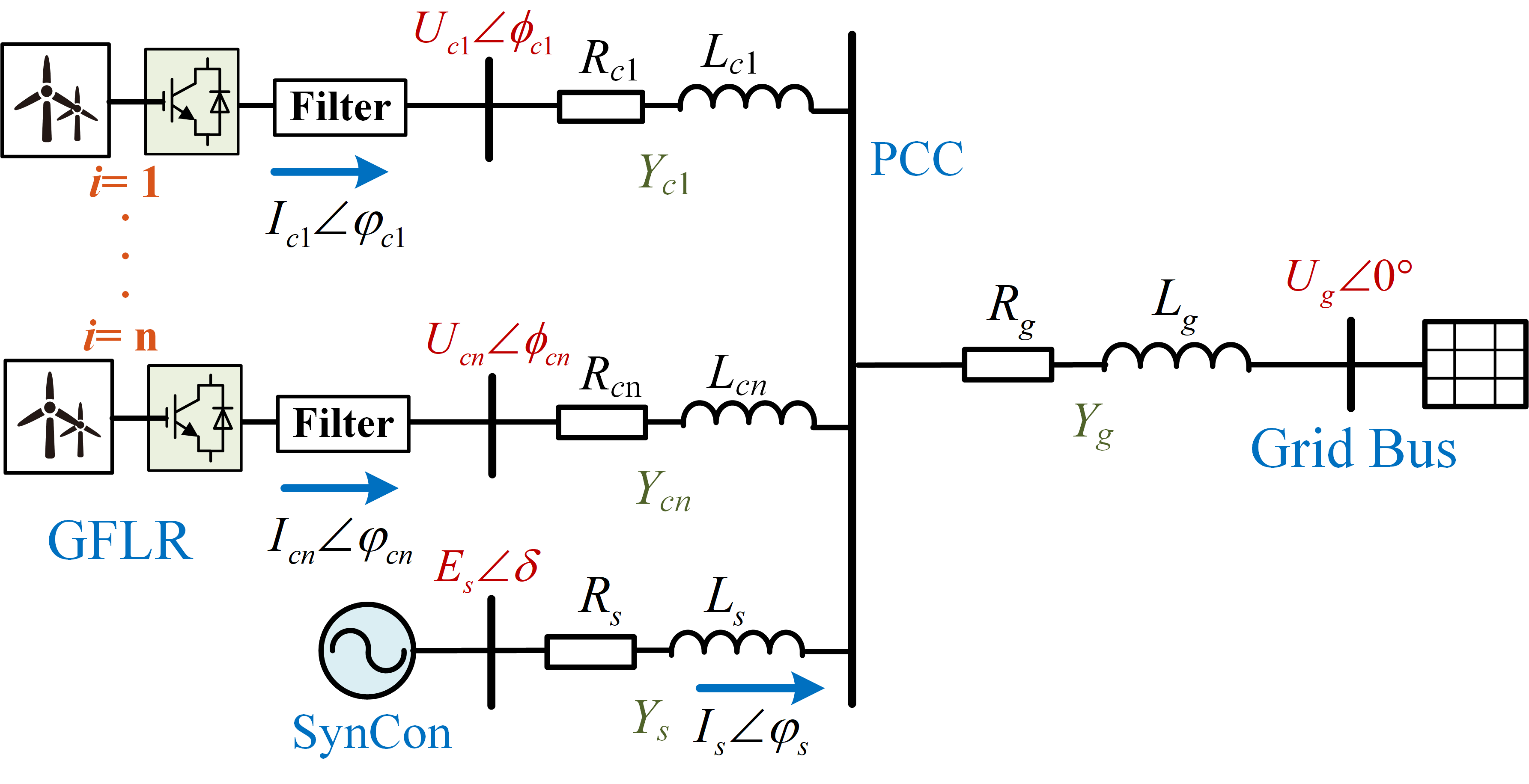}
		\caption{
			\textcolor{black}{Topology of the co-located GFLR and SynCon System.}}
		\label{fig_1}
	\end{figure}

	As shown in Fig.~\ref{fig_1}, \textcolor{black}{the studied power system consists of GFLCs and a SynCon connected at the point of common coupling (PCC)}. The PCC is linked to the remote grid bus via transformers and transmission lines. \textcolor{black}{The co-located system} includes multiple converters, which, after aggregation at the same bus and subsequent aggregation of coherent converters across different buses, can be represented by $n$ aggregated parallel converter units\cite{32}.
    
    The electric potential of the SynCon is denoted by 
	$E_s\angle\delta$. The voltages at the outlet bus of the $i$-th aggregated converter unit (hereafter referred to as the $i$-th converter), and the grid bus are represented by 
	$U_{ci} \angle \phi_{ci} \ (i=1,2,\dots,n)$ and 
	$U_g \angle 0^{\circ}$, respectively. The angular speeds of the PLL of the $i$-th converter, SynCon's rotor, and the grid bus are denoted as {$\omega_{ci}$}, $\omega_s$, and $\omega_g$.
	$I_{ci}\angle\varphi_{ci}$ and $I_s\angle\varphi_s$ are the brunch currents. $R_{ci}$, $R_s$, $R_g$ are the branch resistances, and 
	{$L_{ci}$}, $L_s$, $L_g$ are the branch inductance. The admittance of these branches are denoted as {$Y_{ci}$}, $Y_s$, and $Y_g$.
	\begin{figure}[!t]
		\centering
		\includegraphics[width=2.2in]{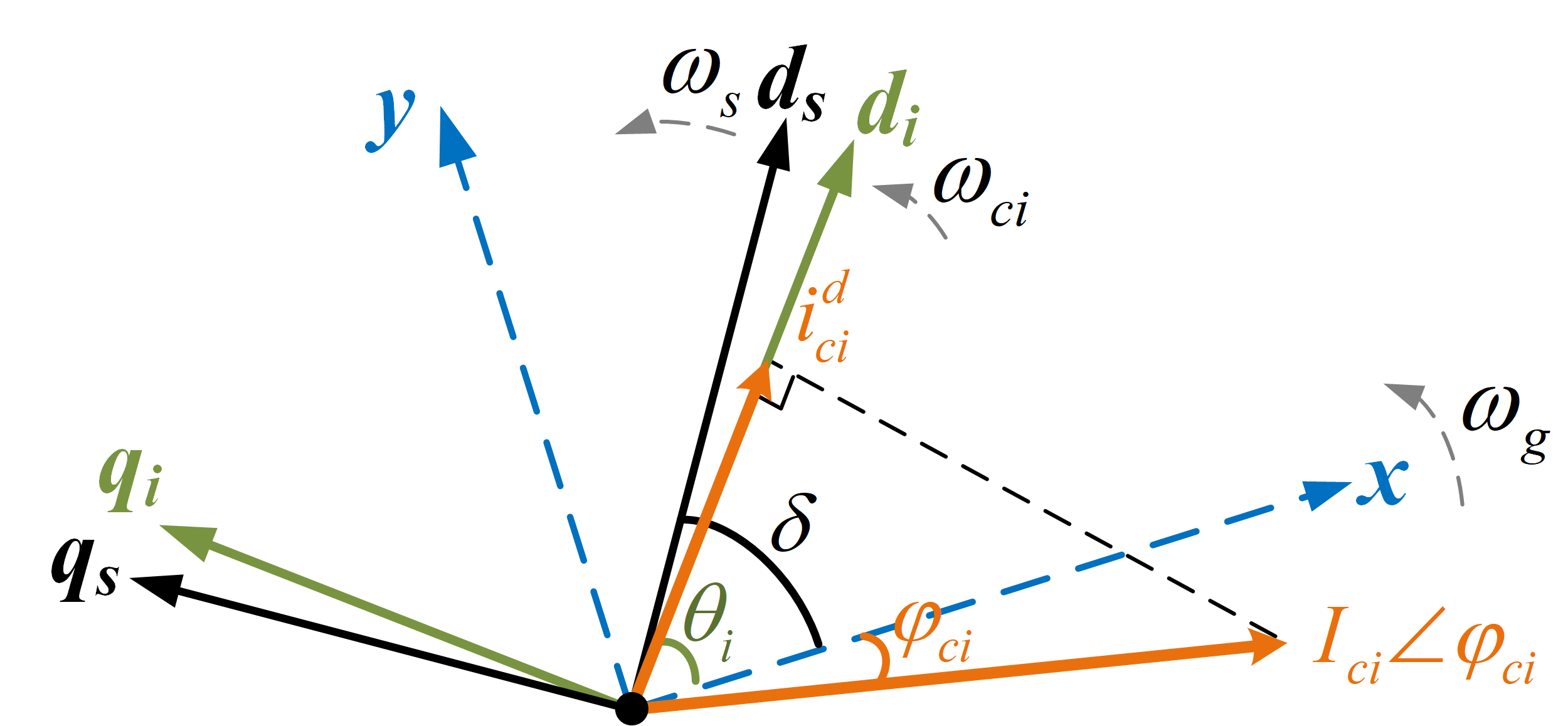}
		\caption{Relationships of different frames in vector space.}
		\label{fig_2}
	\end{figure}
	
	As depicted in Fig.~\ref{fig_2}, \textcolor{black}{the $d_i\!-\!q_i$, $d_s\!-\!q_s$, and $x\!-\!y$ reference frames are the rotating reference frames of the $i$-th converter, the SynCon, and the grid bus, respectively}. These frames rotate counterclockwise at angular speeds of  $\omega_{ci}$, $\omega_s$, and $\omega_g$, respectively.
	The $d_s$-axis leads the $x$-axis by an angle $\delta$, representing the rotor angle of the SynCon relative to the grid bus. 
	The $d_i$-axis leads the $x$-axis by an angle $\theta_i$. 

\begin{figure}[!t]
		\centering
		\includegraphics[width=3.3in]{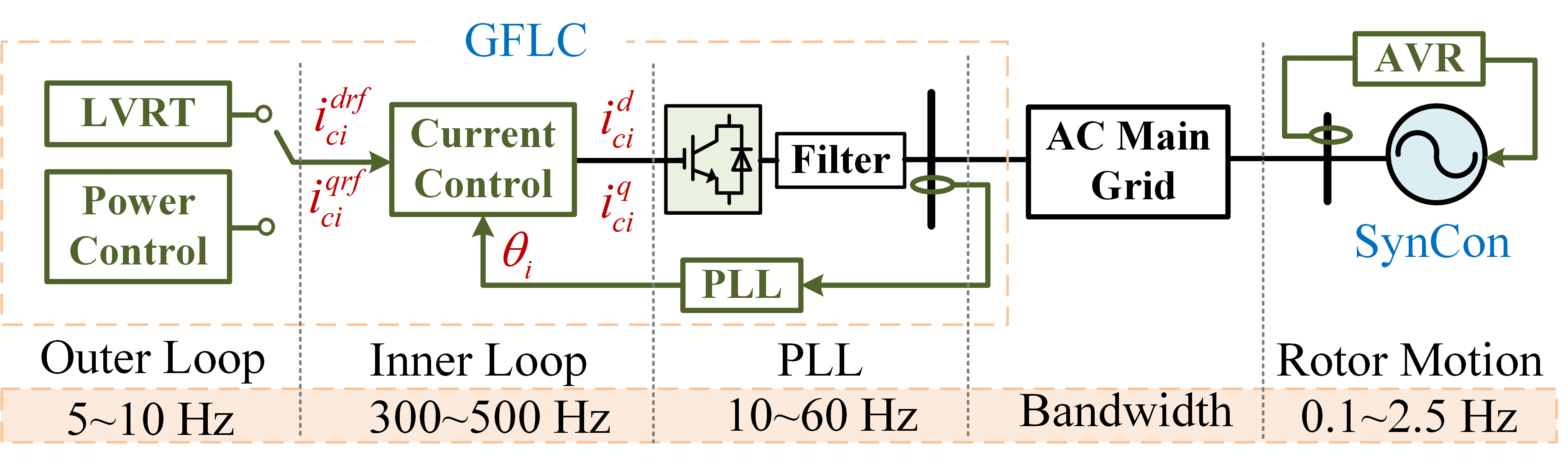}
		\caption{
			{Control structure and time scale division of the GFLR-SynCon system.}}
		\label{fig_3}
	\end{figure}

	The $i-th$ GFLC adopts the classic cascaded control structure shown in Fig.~\ref{fig_3}. 
	Under normal operation, the references for the active and reactive currents, 
	$i_{ci}^{drf}$ and $i_{ci}^{qrf}$, are determined by the outer control loop. 
	During LVRT and its recovery period, these references are directly set by the LVRT control strategy. 
	Since variations in $i_{ci}^{drf}$ and $i_{ci}^{qrf}$ are approximately decoupled from the PLL dynamics, 
	they are treated as parameters rather than state variables in the PLL stability analysis~\cite{29,37}.  
	\textcolor{black}{Furthermore, the inner current control loop exhibits significantly faster dynamics (hundreds of Hz) compared to the PLL dynamics (tens of Hz)~\cite{29}. Thus, when analyzing PLL transient stability, it is reasonable to assume that GFLC currents accurately track their references, i.e., $i_{ci}^d = i_{ci}^{drf}$ and $i_{ci}^q = i_{ci}^{qrf}$~\cite{29,31}}.  Here, $i_{ci}^d$ and $i_{ci}^q$ denote the active and reactive currents of the $i$-th converter, 
	which also satisfy $i_{ci}^d = I_{ci} \cos(\theta_i - \varphi_{ci})$ and 
	$i_{ci}^q = -I_{ci} \sin(\theta_i - \varphi_{ci})$. These relationships are shown in Fig.~\ref{fig_2}.

	\subsection{Dynamic Model of SynCon}
	Without a prime mover, SynCon's real mechanical power is approximately zero. Its rotor dynamic equation is
\begin{equation}
	\label{eq1-2}
	\left\{
	\begin{aligned}
		\frac{d\delta}{dt} &= \omega_s - \omega_g = \omega_g \,\Delta\omega, \\[3pt]
		\frac{d\Delta\omega}{dt} &= \frac{1}{T_s} \left(0 - P_E - D_s \,\Delta\omega \right),
	\end{aligned}
	\right.
\end{equation}
	where \textcolor{black}{$P_E$, $T_s$, and $D_s$ are the mechanical power, electromagnetic power, inertia time constant, and damping coefficient of SynCon, respectively}. 

The electromagnetic power of the SynCon, $P_E$, can be derived from its definition as:
	\begin{equation}	
	\label{eq1-6}
	\begin{aligned}
		P_E&= \mathop{\rm Re}\nolimits \left( {{E_s}\angle \delta  \cdot {I_s}\angle  - {\varphi _s}} \right)\\
		&= \underbrace{{E_s}{U_g}\alpha\left| {{Y_{g}}} \right|\sin \delta}_{P_{es}}  - \underbrace{\alpha {E_s}\sum_{i=1}^n{I_{ci}}\cos \left( {\delta  - {\varphi_{ci}}} \right)}_{P_c},
	\end{aligned}
	\end{equation}
	where $P_{c}$ denotes the power coupling term between the GFLR and the SynCon; \textcolor{black}{$P_{es}$ represents the electromagnetic power of the SynCon in the absence of any influence from the GFLR}; \textcolor{black}{$\alpha$ is a coupling coefficient}, which depends solely on the network topology parameters:
	\begin{equation}
		\label{eq1-5}
		\alpha=\frac{Y_s}{Y_s + Y_g}.
	\end{equation}
	Neglecting the network resistance\cite{37,38} allows $\alpha$ to be a real number. 
	\subsection{Dynamic Model of PLLs}
	The dynamic equations of the $i$-th converter's PLL (hereafter referred to as the $i$-th PLL) are provided as:
\begin{equation}
	\label{eq2}
	\left\{
	\begin{aligned}
		\frac{d\theta_i}{dt} &= \omega_{ci} - \omega_g = \varpi_i, \\[3pt]
		\frac{d\varpi_i}{dt} &= k_{Ii} u_{ci}^{q} + k_{Pi} \frac{d u_{ci}^{q}}{dt},
	\end{aligned}
	\right.
\end{equation}
	where \textcolor{black}{$\varpi_i = \omega_{ci} - \omega_g$ denotes the deviation of the output angular frequency 
	of the $i$-th PLL from the synchronous angular frequency}. 
	$u_{ci}^{q}$ is the $q_i$-axis component of $U_{ci} \angle \phi_{ci}$. $k_{Ii}$ and $k_{Pi}$ are the integral and proportional gains of the $i$-th PLL, respectively.
	
	After the grid connection of the SynCon, $u_{ci}^q$ can be expressed as:
\begin{equation}
	\label{eq3-4}
	\begin{aligned}
		u_{ci}^q &= -\alpha E_s \sin(\theta_i - \delta) - (1-\alpha) U_g \sin\theta_i \\
		&\quad + (\varpi_i + \omega_g) \left[ L_{ci} + (1-\alpha) L_g \right] i_{ci}^d \\
		&\quad + (\varpi_i + \omega_g) (1-\alpha) L_g i_{cij}^{d_i},
	\end{aligned}
\end{equation}
	where
	\textcolor{black}{$i_{cij}^{d_i}$ is the sum of the $d_i$-axis components
	of the currents from all other converters ($j \neq i$). }
	Its expression is given as:
	\begin{equation}
		i_{cij}^{d_i} = \sum_{\substack{j=1,j \neq i}}^{n} I_{cj} \cos(\theta_i - \varphi_j).
		\label{eq:icij_di}
	\end{equation}
    By substituting \eqref{eq3-4} and its time derivative into \eqref{eq2}, the PLL's dynamic equation after SynCon grid integration is as follows:

	\begin{equation}
		\label{eq3-7}
		\frac{d^2\theta_i}{dt^2} = \frac{d\varpi_i}{dt} 
		= \frac{1}{T_{ci}} \left( P_{Mci} - P_{Eci} - D_{ci} \varpi_i - D_{cij} \right),
	\end{equation}
	where \textcolor{black}{$T_{ci}$, $P_{Mci}$, $P_{Eci}$, $D_{ci}$, and $D_{cij}$ denote 
	the equivalent dynamic time constant, the equivalent mechanical power, equivalent electromagnetic power, 
	equivalent damping coefficient with respect to the synchronous grid, 
	and equivalent damping term with respect to other PLLs of the $i$-th PLL, respectively}. 
	Their expressions are given as follows:
	\begin{equation}
		P_{Eci} = \alpha E_s \sin(\theta_i - \delta) + (1-\alpha) U_g \sin\theta_i,
		\label{eq3-7a}
	\end{equation}
	\begin{equation}
		P_{Mci} = \omega_g \left[ L_{ci} + (1-\alpha) L_g \right] i_{ci}^d 
		+ \omega_g (1-\alpha) L_g i_{cij}^{d_i},
		\label{eq3-7b}
	\end{equation}
	\begin{equation}
		\begin{aligned}
			D_{ci} &= \frac{k_{Pi}}{k_{Ii}} \left[ \alpha E_s \cos(\theta_i - \delta) 
			+ (1-\alpha) U_g \cos\theta_i \right] \\
			&\quad - \left[ L_{ci} + (1-\alpha) L_g \right] i_{ci}^d 
			- (1-\alpha) L_g i_{cij}^{d_i},
		\end{aligned}
		\label{eq3-7c}
	\end{equation}
	\begin{equation}
		D_{cij} = \frac{k_{Pi}}{k_{Ii}} \, \omega_g (1-\alpha) L_g \, \sum_{\substack{j=1,j \neq i}}^{n} I_{cj} \sin(\theta_i - \varphi_j) (\varpi_i - \varpi_j),
		\label{eq3-7d}
	\end{equation}
	\begin{equation}
		T_{ci} = \frac{1}{k_{Ii}} \left\{ 1 - k_{Pi} \left[ \left( L_{ci} + (1-\alpha) L_g \right) i_{ci}^d 
		+ (1-\alpha) L_g i_{cij}^{d_i} \right] \right\}.
		\label{eq3-7e}
	\end{equation}
	
	\subsection{Singular Perturbation Model of the co-located System}
    Due to the significant difference in time scales between the rotor dynamics of the SynCon 
    and the control response of the PLLs\cite{34}, their overall dynamic behavior exhibits a dual time-scale characteristic. 
    By introducing the dimensionless parameter:
    \begin{equation}
    	\varepsilon =\frac{T_{ci}}{T_s}=\frac{1 - k_{Pi} \left[  (L_{ci} + (1-\alpha) L_g) i_{ci}^d + (1-\alpha) L_g i_{cij}^{d_i} \right]}{k_{Ii}T_{s}},
    	\label{eq_eps}
    \end{equation} 
    the original model of the co-located GFLR and SynCon system can be decoupled into two subsystems. 
    They are a slow subsystem associated with the rotor dynamics of the SynCon, 
    and a fast subsystem associated with the dynamics of the PLL:
    \begin{equation}
    	\label{eq3-11}
    	\text{Slow subsystem:} 
    	\left\{
    	\begin{aligned}
    		&\frac{d\delta}{dt} = \Delta \omega, \\[3pt]
    		&T_s \frac{d\Delta \omega}{dt} = P_{c} - P_{es} - D_s \Delta \omega, \\[3pt]
    		&0 = P_{Mci} - P_{Eci}.
    	\end{aligned}
    	\right.
    \end{equation}
    \begin{equation}
    	\label{eq3-12}
    	\hspace{-3.5mm}
    	\text{Fast subsystem:} 
    	\left\{
    	\begin{aligned}
    		&\frac{d\delta}{d\tau} = 0, \frac{d\Delta \omega}{d\tau} = 0, \\[3pt]
    		&\frac{d\theta_i}{d\tau} = \varepsilon \varpi_i, \\[3pt]
    		&\frac{d\varpi_i}{d\tau} = \frac{1}{T_s}(P_{Mci} - P_{Eci} - D_{ci} \varpi_i - D_{cij}).
    	\end{aligned}
    	\right.
    \end{equation}
	where $\tau=t/\varepsilon$. \textcolor{black}{Based on singular perturbation theory\cite{39}, when analyzing the fast subsystem’s stability, the slow subsystem’s state variables are considered “frozen” (or quasi-steady). Conversely, when investigating the stability of the slow subsystem, the fast subsystem is assumed to have reached its steady state (or to have become unstable).}

	\section{\textcolor{black}{Mechanism of Instability Source Transfer Induced by SynCon Integration}}
 
    \textcolor{black}{This section elucidates the impact mechanism of SynCon integration on system stability by analyzing its role in fast-time-scale stability and its own slow-time-scale dynamics.  Consequently, it reveals how the dominant instability source transitions from the fast to the slow time scale.}
	
	\subsection{\textcolor{black}{SynCon's Role in Enhancing Fast Time-Scale Coherent PLLs Stability}}
	Let the superscript ``$*$'' denote the value of a variable of the $i$-th converter 
	under the condition that all aggregated parallel converter units operate coherently. 
	For example, \textcolor{black}{
	$i_{cij}^{d_i*}$ represents the $d_i$-axis component of the currents 
	from other converters ($j \neq i$) under coherent condition. 
	Accordingly, $P_{Mci}^*$ and $D_{ci}^*$ represent, respectively, 
	the equivalent mechanical power and the damping coefficient with respect to the synchronous grid, both under coherent condition}. 
	The coherent-operation case is regarded as the nominal system of the $i$-th converter, 
	as given in~\eqref{n1}:
	\begin{equation}
		\label{n1}
		\frac{d^2\theta_i}{dt^2} = \frac{d\varpi_i}{dt} 
		= \frac{1}{T_{ci}} f_i\left(\theta_i, \overline{\delta}\right),
	\end{equation}
	where $\overline{\delta}$ is the frozen value of $\delta$ in the fast time scale, and
	\begin{equation}
		\label{n2}
		f\left(\theta_i, \overline{\delta}\right) 
		= P_{Mci}^* - P_{Eci} - D_{ci}^* \varpi_i.
	\end{equation}	
	
	\textcolor{black}{The angle at the stable equilibrium point (SEP) of the $i$-th PLL, denoted by $\theta_{si}^{*}$}, can be expressed as:
	\begin{equation}
		\label{sep}
		\theta_{si}^{*} = \arcsin\!\left(\frac{P_{Mci}^{*}}{U_E}\right)
		+ \underbrace{\arctan\!\left(\frac{\alpha E_s\sin\overline{\delta}}{\alpha E_s\cos\overline{\delta}+(1-\alpha) U_g}\right)}_{\theta_{\delta}},
	\end{equation}
	where \textcolor{black}{the arctan term of $\theta_{si}^{*}$ is denoted as $\theta_{\delta}$}, and the expression for $U_E$ is given by:
\begin{equation}
	\label{UE}
	U_E = \sqrt{\alpha^2 E_s^2 + (1-\alpha)^2 U_g^2 + 2\alpha(1-\alpha)E_s U_g\cos\overline{\delta}}.
\end{equation}
\textcolor{black}{It follows that $U_E$ can be regarded as an equivalent voltage established after the SynCon is connected to the grid. If $\alpha = 0$ (i.e., SynCon not integrated), $U_E = U_g$.}

\textcolor{black}{Previous studies have shown that the range of $\theta$ for which the damping coefficient is positive defines the PLL’s conservative stability boundary\cite{29,42}.  
For $D_{ci}^*>0$, the conservative stability boundary of the $i$-th PLL is given by:}
	\begin{equation}
		\label{boundary}
		\theta_{\delta} - \theta_{bi} < \theta_i < \theta_{\delta} + \theta_{bi},
	\end{equation}
	where
\begin{equation}
	\label{theta_b}
	\theta_{bi} = \arccos\!\left(\frac{k_{Ii}P_{Mci}^*}{k_{Pi}\omega_gU_E}\right). 
\end{equation}
	
\textcolor{black}{By applying a coordinate transformation $\theta_i^{'} = \theta_i - \theta_{\delta}$,
the original boundary in~\eqref{boundary} is shifted to the  symmetric form $-\theta_{bi} < \theta_i^{'} < \theta_{bi}$. This translation does not affect the system’s intrinsic dynamics, because the governing differential equations remain invariant under constant shifts of state variables\cite{41}.
	In this transformed system, $\theta_{bi}$ in directly represents the half-width of the positive-damping region. 	As discussed in~\cite{42}, an enlargement of the positive-damping region leads to an enhancement of the actual stability boundary.  
	We therefore adopt $\theta_{bi}$ as the stability metric for the $i$-th PLL.}
 
	
\textcolor{black}{This metric (eq.\eqref{theta_b}) shows that the integration of the SynCon changes the transient stability boundary of PLLs by adjusting both $P_{Mci}^*$ and $U_E$. A smaller $P_{Mci}^*$ or a larger $U_E$ will expand this stability boundary. We first examine $U_E$. From the voltage phasor relationship, it is evident that in steady state, the per-unit value of $U_E$ is greater than $U_g$. }
During a fault, eq.\eqref{UE} shows that even if $U_g$ drops to zero, $U_E=\alpha E_s$, which means the SynCon's internal voltage can still provide a reference for PLLs.
After the fault is cleared, eq.\eqref{UE} indicates that an increase in $\delta$ reduces $U_E$. However, in practice, protective relays operate rapidly, and faults are typically cleared within $100$ ms. As a result, the change in rotor angle between the pre-fault and clearance instants is small.  
Because SynCon and PLL dynamics evolve on different time scales, the PLL stability boundary is determined by $\bar{\delta}$ at the moment of clearance. \textcolor{black}{On the fast time scale of the PLL, $U_E$ is therefore likely to remain above $U_g$, or at least not much lower, which means its negative influence on the stability boundary is limited. }

\textcolor{black}{In contrast, changes in $P_{Mci}^*$ caused by SynCon have a more pronounced impact on the PLL stability boundary after fault clearance. Eq.\eqref{eq1-5} shows that $\alpha$ is $0$ before the SynCon is connected to the grid. After connection, it increases to a value close to 1 because $Y_s>Y_g$. Thus, $(1-\alpha)L_g$ decreases sharply after the SynCon is integrated. Given that $L_g>>L_{ci}$, this reduction leads to a significant drop in $P_{Mci}^*$, thereby greatly improving PLL stability.}

In summary, \textbf{it is the SynCon's inherent voltage source characteristics and its dynamic time-scale decoupling from the PLL that enable it to provide a robust and proximate voltage reference, thereby significantly expanding the PLL's stability boundary.}  
	
\subsection{\textcolor{black}{SynCon's Role in Mitigating Multi-Converter Non-Coherent Disturbances}}
	\begin{figure}[!t]
		\centering
		\includegraphics[width=3.5in]{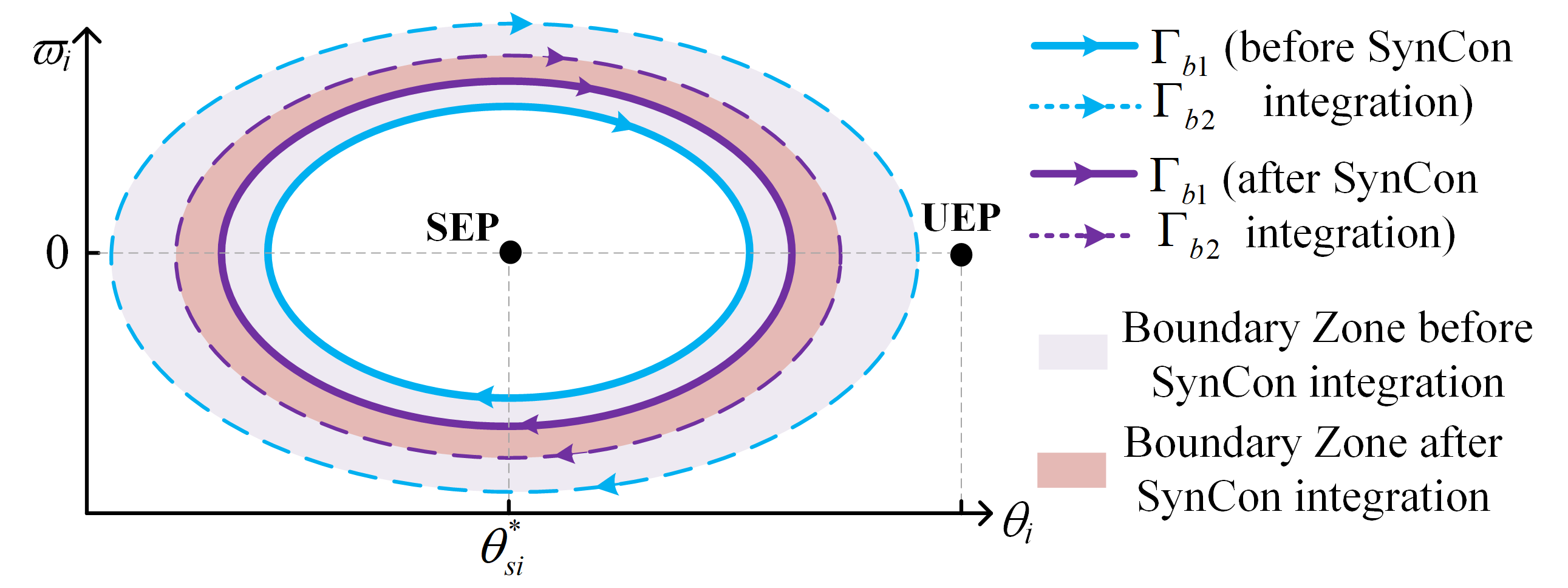}
		\caption{{Uncertainty boundary zone and conservative stability boundary of the perturbed system.}}
		\label{fig_5}
	\end{figure}
	If non-coherent behavior exists among aggregated parallel converter units during transient periods, it introduces perturbations to the nominal system. 
	The dynamic equation of the perturbed system can be expressed as:
	\begin{equation}
		\label{perturbation}
		\frac{d^2\theta_i}{dt^2} = \frac{d\varpi_i}{dt} 
		= \frac{1}{T_{ci}} \left[f_i\left(\theta_i, \overline{\delta}\right)+u_i(\theta_1,...,\theta_n,\overline{\delta})\right],
	\end{equation}
	where $u_i(\theta_1,...,\theta_n,\overline{\delta})$ denotes the  disturbance term of the $i$-th converter. The expression of $u_i$ is:
	\begin{equation}
	\hspace{-2mm}
	\label{ui}
	u_i(\theta_1,...,\theta_n,\overline{\delta})=\Delta P_{Mci}-\Delta D_{ci}\varpi_i-D_{cmi},
	\end{equation}
	where $\Delta P_{Mci}=P_{Mci}-P_{Mci}^*$ and $\Delta D_{ci}=D_{ci}-D_{ci}^*$.
	At the SEP ($\theta_{si}^*, 0$), $u_i$ equals zero. 
		
	Let $u_{i}^{\rm{max}}$ and $u_{i}^{\rm{min}}$ denote the upper and lower bounds of $u_i$, which can be expressed as:
	\begin{equation}
		\label{eq:u_imax}
		\begin{aligned}
			u_{i}^{\rm{max}} = & \left( {{\omega _g} + {\varpi _i}} \right)\Big[{(1 - \alpha ){L_g}\left( {\sum\limits_{j = 1,j\neq i}^n {{I_{cj}^{\max}}}  - i_{cij}^{d_i*} } \right)} \Big]>0,
		\end{aligned}
	\end{equation}
	
\begin{equation}
	\label{eq:u_imin}
	\begin{aligned}
		u_i^{\min } = & -\left( {{\omega _g} + {\varpi _i}} \right)\Big[{(1 - \alpha ){L_g} i_{cij}^{d_i*} } \Big] \\
		& - \frac{{{k_{Pi}}}}{{{k_{Ii}}}}{\mkern 1mu} {\omega _g}(1 - \alpha ){L_g}{\mkern 1mu} \sum\limits_{j \ne i} {({\varpi _i} - {\varpi _j})I_{cj}^{\max }}<0,
	\end{aligned}
\end{equation}
	where \textcolor{black}{$I_{cj}^{\rm{max}}$ corresponds to the maximum allowable current for the $j$-th converters.}
	
	To investigate the effect of disturbance $u_i$ on the nominal system \eqref{n1}, we first define the transient energy function of the system as:
   \begin{equation}
		\label{Lfun}
		\begin{aligned}
			V_{ci}\left( {\varpi_i,\theta_i} \right) = \frac{1}{2}T_{ci}{\varpi_i^2} + \int_{{\theta _{si}^{*}}}^{\theta_i}  {\left( {P_{Eci}-P_{Mci}^{*}} \right)d\theta_i}.
		\end{aligned}
	\end{equation}
	
	{Then, the derivative of $V_{ci}$ along the trajectory of the perturbed system \eqref{perturbation} is given by:}
	\begin{equation}
	\label{n6}
	\begin{aligned}
		\dot V_{ci}=
		-D_{ci}^*{\varpi_i^2}+u_i\varpi_i.
	\end{aligned}
	\end{equation}	
	
	{According to \eqref{n6}, the effect of the disturbance term $u_i$ depends critically on the sign of $u_i\varpi_i$. If $u_i\varpi_i<0$, $u_i$ contributes to energy dissipation, with a larger $|u_i|$ leading to more significant decay. Conversely, if $u_i\varpi_i>0$, $u_i$ drives an increase in system energy, and a larger $|u_i|$  results in a more pronounced energy rise. }
	
	{Consequently, the switching conditions for system \eqref{n7} are specifically designed to maximize the increase in the system's periodic energy caused by the disturbance term $u_i$. Conversely, the switching conditions for system \eqref{n8} are formulated to maximize the dissipation of this energy. }	
	\begin{equation}
		\label{n7}
		\begin{aligned}
			\frac{d^2\theta_i}{dt^2}=\frac{d\varpi_i}{dt}=\left\{ \begin{array}{l}
				\frac{1}{T_{ci}}(f_i+u_{i}^{\rm{max}}), \quad \varpi_i>0, \\[1.5ex] 
				\frac{1}{T_{ci}}(f_i+u_{i}^{\rm{min}}), \quad \varpi_i<0.\\
			\end{array} \right.
		\end{aligned}
	\end{equation}
	\begin{equation}
		\label{n8}
		\begin{aligned}
			\frac{d^2\theta_i}{dt^2}=\frac{d\varpi_i}{dt}=\left\{ \begin{array}{l}
				\frac{1}{T_{ci}}(f_i+u_{i}^{\rm{min}}), \quad \varpi_i>0, \\[1.5ex] 
				\frac{1}{T_{ci}}(f_i+u_{i}^{\rm{max}}), \quad \varpi_i<0.\\
			\end{array} \right.
		\end{aligned}
	\end{equation}
	
	{Collectively, all possible boundaries of the perturbed system \eqref{perturbation} define an ``uncertainty boundary zone". As shown in Fig.~\ref{fig_5}, the boundaries of switched systems \eqref{n7} and \eqref{n8} correspond to the inner boundary $(\Gamma_{b1})$ and outer boundary $(\Gamma_{b2})$ of this zone, respectively. The nominal system boundary and the actual boundary under perturbation both lie within this zone. Consequently, inner boundary $(\Gamma_{b1})$ represents the conservative stability boundary of the $i$-th converter when subjected to multi-converter non-coherent perturbations.  }
	
	{As shown above, an increase in $\left|u_{i}^{\rm{max}}\right|$ and $\left|u_{i}^{\rm{min}}\right|$ results in greater energy accumulation, which in turn causes the conservative boundary to shrink. Given that $0 < \alpha < 1$, the integration of SynCon effectively reduces the values of $\left|u_{i}^{\rm{max}}\right|$ and $\left|u_{i}^{\rm{min}}\right|$ compared to the scenario without SynCon. This reduction consequently implies less energy accumulation. Thus, \textbf{SynCon mitigates the uncertainty caused by non-coherent behavior of multiple GFLCs and expands the conservative stability boundary of the perturbed system.}  
		
	\textcolor{black}{Furthermore, if a decrease in \(|u_{i}^{\rm{max}}|\) and \(|u_{i}^{\rm{min}}|\) is caused by an increase in \(\alpha\), this would consequently lead to a decrease in \(P_{Mci}^*\). This demonstrates that the metric proposed in \eqref{theta_b} can also characterize the impact of SynCon integration on the conservative stability boundary of the perturbed system.}

	\subsection{\textcolor{black}{Emergence of Slow Time-Scale SynCon Angle Instability}}
	Next, we shift our focus to SynCon's rotor angle stability on the slow time scale.
     Initially, one might assume that its zero mechanical power makes it inherently stable. However, according to \eqref{eq3-11}, 
    the power coupling term $P_c$ effectively functions as  SynCon's ``mechanical power'', which can be expressed as
	\begin{equation}
		\label{6-5}
		\begin{aligned}
			P_c=\alpha {E_s}\sum_{i=1}^n{I_{ci}}\cos \left( {\delta  - {\theta_{i}}+\eta_{ci}}\right)=\alpha {E_s}i_c^{d_s},
		\end{aligned}
	\end{equation}
	 where \textcolor{black}{$\eta_{ci}=\theta_i-\varphi_{ci}$ represents the power factor angle of the $i$-th converter current}. Under steady-state conditions or after fault clearance, $\eta_{ci}$ typically becomes zero. In \eqref{6-5}, the term $\sum_{i=1}^n{I_{ci}}\cos \left( {\delta  - {\theta_{i}}+\eta_{ci}}\right)$ represents the sum of the projections of each GFLC current $I_{ci}\angle\varphi_{ci}$ onto the $d_s$-axis. This projection, denoted as $i_c^{d_s}$, is then multiplied by the coefficient $\alpha E_s$ to yield $P_c$.
	 
     According to singular perturbation theory, fast subsystem dynamics can be neglected when analyzing slow subsystem behavior. Thus, under GFLR's grid-following control, stable PLLs can quickly track frequency variations caused by SynCon's angle oscillations, implying that $\delta$ and
     $\theta$ tend to vary in the same direction. Given the close electrical coupling between GFLR and SynCon, we can further approximate $\delta\approx\theta_i$. Consequently, 
     as evident from \eqref{6-5}, PLLs' ability to follow the grid ensures that $P_c$ is mainly governed by the GFLR's active current.  
     \textcolor{black}{It is due to the presence of this equivalent mechanical power that the SynCon can accumulate transient energy during faults\cite{27}. If the system fails to dissipate this accumulated energy after fault clearance, the SynCon faces a risk of transient instability. Therefore, $P_c$ in \eqref{6-5} can be regarded as a transient stability indicator for the SynCon. An increase in $P_c$ signifies a deterioration in SynCon's stability.}

  	\begin{figure}[!t]
  		\centering
  		\includegraphics[width=3.4in]{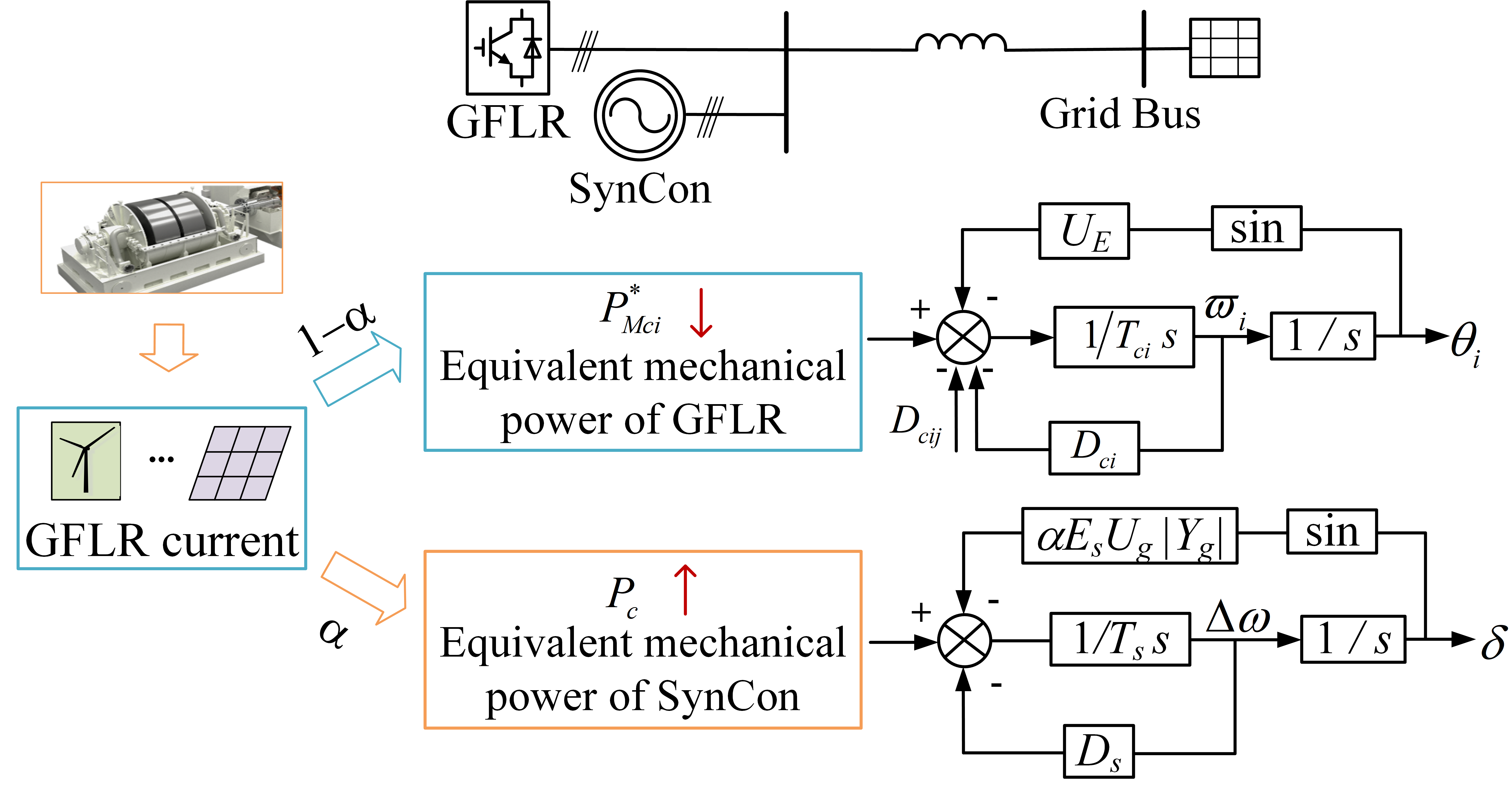}
  		\caption{\textcolor{black}{Comparison of equivalent dynamic systems of PLL and SynCon.}}
  		\label{fig_9}
  	\end{figure}

    \subsection{\textcolor{black}{The Instability Source Transfer: From Fast to Slow Time Scales}}
	Synthesizing the dual-time-scale stability analysis from subsections A, B, and C in Section III, it is revealed that the stability of both fast and slow time scales is primarily determined by the equivalent mechanical power of their respective dynamic systems, specifically $P_{Mci}^*$ and $P_c$. Eqs. \eqref{PMci_with_syncon_change} and \eqref{Pc_change} show how they change with SynCon integration:
\begin{equation}
	\label{PMci_with_syncon_change} 
	\begin{split}
		P_{Mci}^* &= \omega_g \left[ (L_{ci} + L_g) i_{ci}^{d} + L_g i_{cij}^{d_i*} \right]\xrightarrow{\text{SynCon integration}} \\
		P_{Mci}^* &= (1-\alpha)\omega_g \left( \gamma_i i_{ci}^{d} + L_g i_{cij}^{d_i*} \right),
	\end{split}
\end{equation}
	\begin{equation}
		\label{Pc_change} 
		P_c = 0 \xrightarrow{\text{SynCon integration}} P_c = \alpha E_s i_c^{d_s},
	\end{equation}
	where $\gamma_i=({L_{ci}}{L_g} + {L_{ci}}{L_s} + {L_s}{L_g})/{L_s}$. A comparison of \eqref{PMci_with_syncon_change} and \eqref{Pc_change} reveals that the GFLRs' current (particularly its active current) serves as a common driving force for instability in both the PLLs and the SynCon. This driving force is distributed between the PLLs and the SynCon with ``distribution coefficients" of $1-\alpha$ and $\alpha$, respectively, as shown in Fig.~\ref{fig_9}. When SynCon is not integrated, $\alpha=0$, and the entire GFLR current  contributes to the PLLs' equivalent mechanical power. Upon SynCon integration, due to the tight electrical coupling between SynCon and GFLR, $\alpha$ takes a value such that 
	close to $1$.  This implies that the SynCon bears a greater share of the instability risk. This reallocation of the destabilizing force signifies a shift in the system's dominant instability concern from the fast time-scale PLLs to the slow time-scale SynCon.
	
	\begin{figure}[!t]
		\centering
		\includegraphics[width=3.6in]{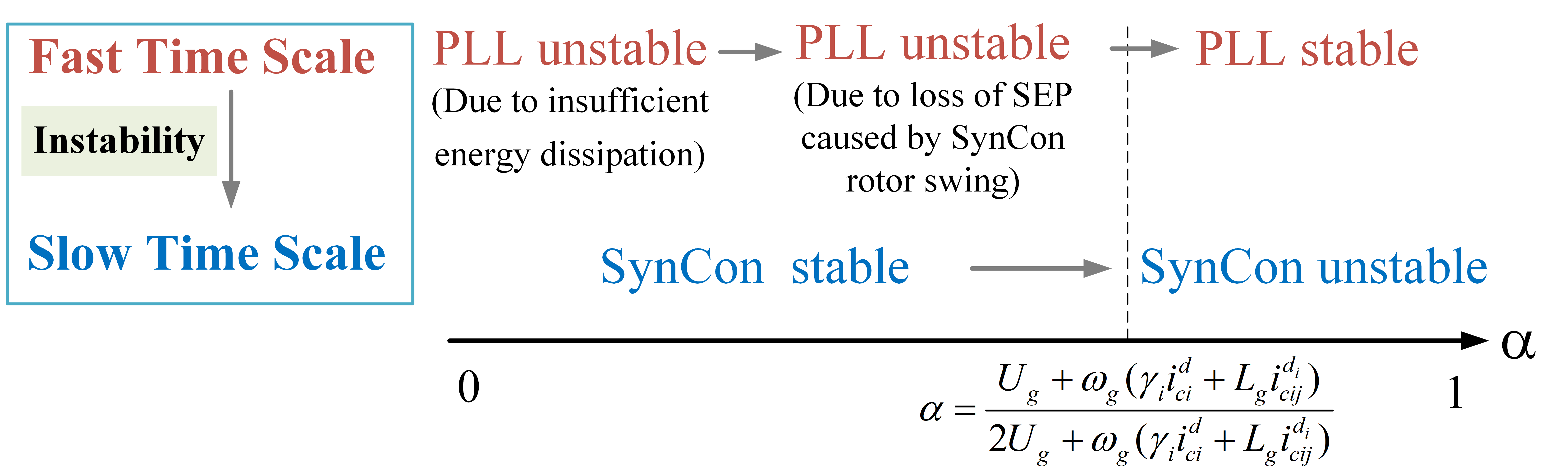}
		\caption{Transition of instability source of the co-located system.}
		\label{fig_17}
	\end{figure}

    However, the increase in SynCon's rotor angle may cause PLLs to lose SEPs\cite{34}. If the GFLR loses its grid-following capability or disconnects, $P_c$ will be reduced or disappear. As a result, SynCon is more likely to stabilize because the effective ``mechanical power" drops to zero midway. 
    Considering this risk, does the established pattern of dominant instability source transfer remain applicable? 
    
    After PLLs convergence, the $i$-th PLL can be simplified to an algebraic equation in the slow time scale:
	\begin{equation}
	\label{n12}
	\begin{aligned}
		U_E\sin \left({\theta_i-\theta_\delta} \right)
		-{\omega _g}(1-\alpha)({\gamma_i}i_{ci}^d+L_g{i_{cij}^{d_i}})=0.
	\end{aligned}
	\end{equation}
    The existence of a SEP for the PLL implies that \eqref{n12} has a solution, which is equivalent to:
    \begin{equation}
    	\label{2}
    	\begin{aligned}
    	U_E\ge {\omega _g}(1-\alpha)({\gamma_i}i_{ci}^d+L_g{i_{cij}^{d_i}}).
    	\end{aligned}
    \end{equation}
    
    Eq.\eqref{UE} shows that $U_E$ reaches its minimum when $\delta = \pi + 2k\pi$ (where $k$ is an integer). Inserting this condition into \eqref{2} yields the sufficient condition for the $i$-th PLL to have a SEP:
    \begin{align}
	\label{3}
	\begin{aligned}
	\alpha\geq\frac{U_g+\omega_g({\gamma_i}i_{ci}^d+L_g{i_{cij}^{d_i}})}{2U_g+\omega_g({\gamma_i}i_{ci}^d+L_g{i_{cij}^{d_i}})}\in(0,1).
	\end{aligned}
	\end{align} 

	Eq.\eqref{3} indicates that as $\alpha$ exceeds a certain threshold, PLLs can preserve SEP existence despite SynCon's angle oscillations. 
	
	In summary,\textbf{ as the electrical coupling between GFLR and SynCon strengthens, the primary instability source of the system shifts from fast-timescale PLLs to slow-timescale SynCon}. This process is shown in Fig.~\ref{fig_17}.
	
	\section{Dual-Timescale Stability Enhancement Based on Classical PLL Damping Control}
	Previous analyses suggest that while deploying SynCon enhances GFLR's transient stability, it does not eliminate the risk of synchronization stability in the system but rather transfers this risk to the SynCon itself. 
	To address this issue, \textcolor{black}{this section proposes a simple method that leverages the inherent interaction between SynCons and PLLs across fast and slow time scales. Specifically, it demonstrates that the damping effect provided by a well-tuned classical PLL damping control (as shown in Fig.~\ref{fig_10}) can be effectively transferred from the fast to the slow time scale to damp SynCon rotor acceleration.}   

	\subsection{{The Damping Control effect Transfer: From Fast to Slow Time Scales}}
	{\textbf{1) Damping Effect on the Fast Subsystem}}
	
	The transient damping control for the PLL shown in Fig.~\ref{fig_10} offers a simple yet effective solution for PLL stabilization, where   \textcolor{black}{$K_d$ is the damping loop coefficient}.
	Applying this control updates the dynamic equation of the $i$-th PLL to:
	\begin{equation}
		\label{n13}
		\left\{
		\begin{aligned}
			\frac{{d\theta_i}}{{d\tau}} &=\varepsilon\varpi_i, \\[3pt]
			\frac{{d\varpi_i}}{{d\tau}} &= \frac{1}{T_s}[f_i\left(\theta_i, \overline{\delta}\right)+u_i(\theta_1,...,\theta_n,\overline{\delta})-K_d\varpi_i].
		\end{aligned}
		\right.
	\end{equation}

	{Comparing \eqref{perturbation} and \eqref{n13} reveals that the PLL's equivalent damping coefficient increases from \(D_{ci}\) to \(D_{ci}+K_d\) under this damping control, thereby significantly improving its convergence. Notably, this damping control effect is further enhanced by the reduction in the non-coherent coupling term \(u_i\) resulting from the SynCon integration.}
	
	{\textbf{2) Damping Effect on the Slow Subsystem}}

	{Equation \eqref{6-5} indicates that \(P_c\) reaches its maximum when \(\delta\approx\theta_i\) and \(\eta_i=0\), signifying the most critical condition for SynCon transient stability. Consequently, our subsequent analysis will specifically examine scenarios where the electrical distance between the SynCon and GFLR is small, and the GFLR injects pure active current.
	
	On the slow time scale, \(t_0\) marks the instant after PLL synchronization. We define \(\Delta t\) as a small time interval, during which \(\delta\) and \(\theta_i\) change by \(\Delta\delta\) and \(\Delta\theta_i\), respectively. Let \(t_1\) denote a specific instant  during \(\delta\) acceleration in the first swing after \(t_0\). At this moment, the SynCon's ``equivalent mechanical power" can be expressed as:}
	\begin{equation}
		\label{n23}
		\hspace{-1.mm}
		P_{c}(t_1)=\alpha {E_s}\sum_{i=1}^n{I_{ci}}\cos \left[{\delta_0-\theta_{i0}+\int{(\Delta\delta(\Delta t)-\Delta\theta_i(\Delta t))}} \right],
	\end{equation}
	{where $\delta_0$ and $\theta_{i0}$ represent the value of $\delta$ and $\theta_i$ at $t_0$, respectively. Since $\Delta\delta$ varies slightly, we perform small-signal linearization of the PLL dynamics under this SynCon-induced disturbance.
	If the SynCon and GFLR are electrically close, $\Delta\delta$ can be treated as the tracking target for $\Delta\theta_i$. Defining $G(s)$ as the transfer function from $\Delta\delta(s)$ to $\Delta\theta_i(s)$, we obtain:  }
	\begin{figure}[!t]
		\centering
		\includegraphics[width=3in]{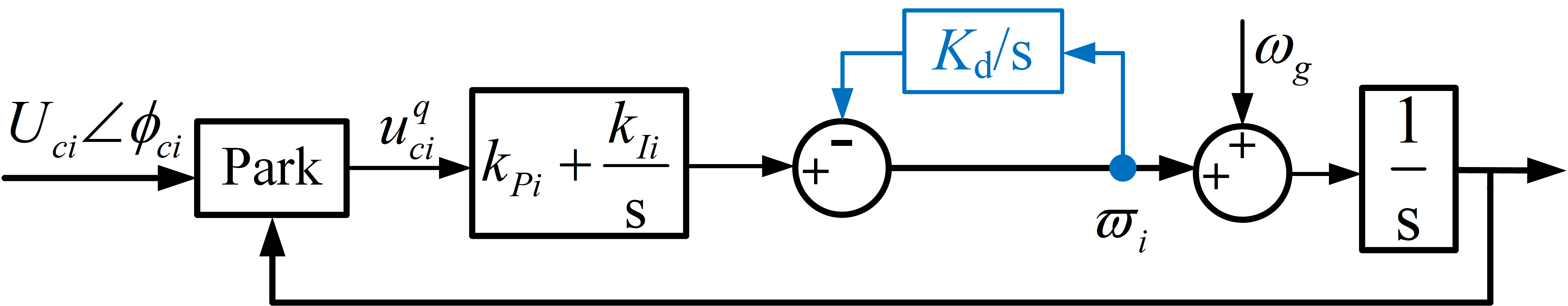}
		\caption{{Scheme of classical PLL damping control}}
		\label{fig_10}
	\end{figure}
	\begin{figure}[!t]
	\centering
	\includegraphics[width=1.6in]{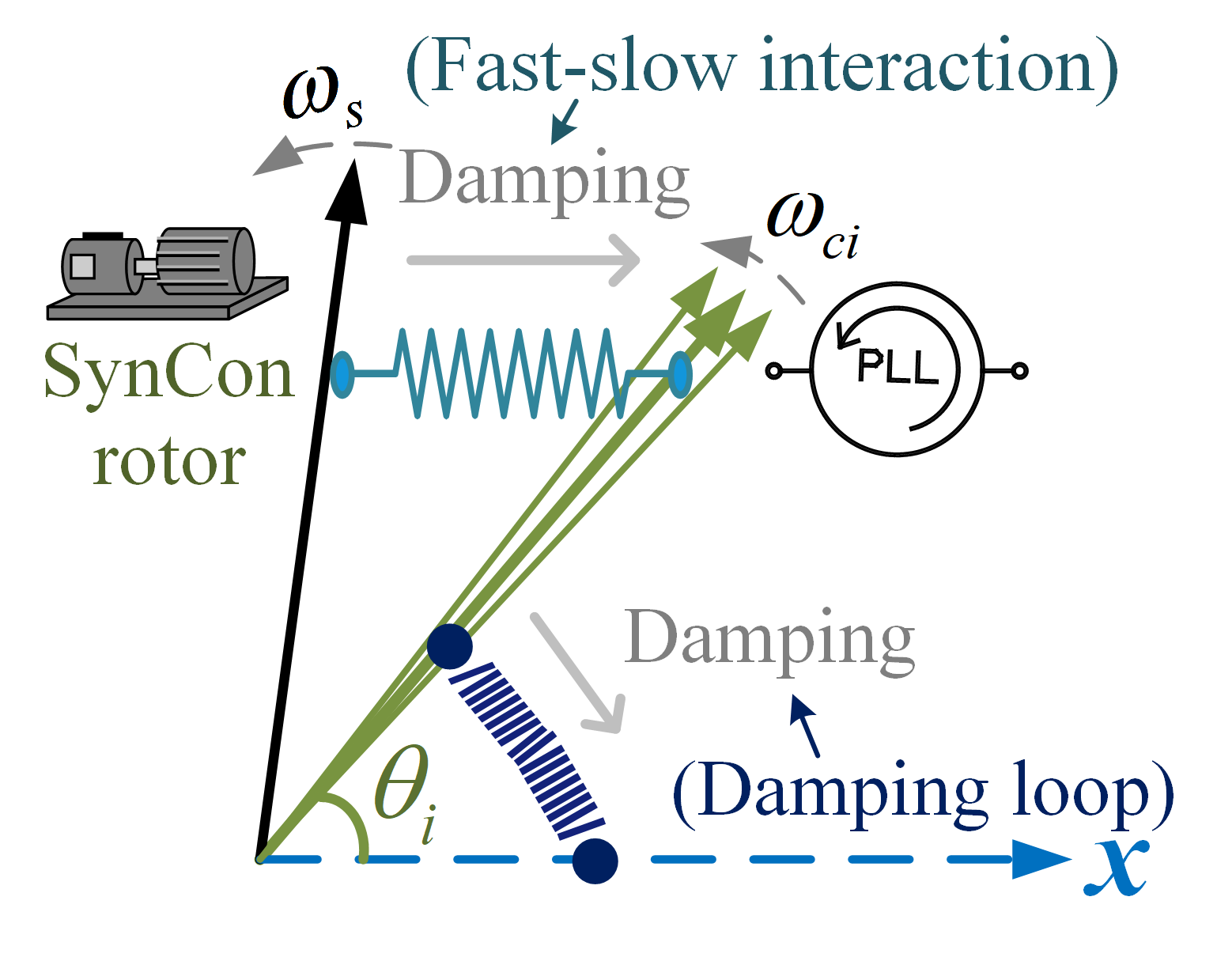}
	\caption{\textcolor{black}{Damping Effect of PLL Control on SynCon.}}
	\label{fig_16}
\end{figure}
	\begin{equation}
		\label{n14}
		\hspace{-1.mm}
		G(s)=\frac{U_{ci}(k_{Pi}s+k_{Ii})}{s^2+(k_{Pi}U_{ci}+K_d)s+k_{Ii}U_{ci}}.
	\end{equation}
	
	{Rearranging the equation $\Delta\theta_i(s)=G(s)\Delta\delta(s)$ yields:}
	\begin{equation}
		\label{n17}
		\hspace{-2.5mm}
		\Delta\delta(s)-\Delta\theta_i(s)=\frac{s^2+K_ds}{s^2+(k_{Pi}U_{ci}+K_d)s+k_{Ii}U_{ci}}\Delta\delta(s).
	\end{equation}
	
	Next, we analyze the relationship between the integral term \(\int{(\Delta\delta-\Delta\theta_i)}\) and the damping coefficient \(K_d\).
	The time-scale separation property ensures that the fast subsystem continuously tracks perturbations from the slow subsystem. This implies that for a given \(\Delta t\) on the slow time scale (typically \(\Delta t > 10\varepsilon\), based on engineering convergence principles), the fast subsystem completes its evolution over \(\tau=\Delta t/\varepsilon\) and subsequently converges. From an engineering perspective, ``infinite time" is interpreted as the duration required for guaranteed system convergence. Consequently, \(\tau\) effectively approaches infinity in this context. This justification allows us to apply the Final Value Theorem to evaluate \(\Delta\delta(\Delta t)-\Delta\theta_i(\Delta t)\):
	\begin{equation}
		\label{n24}
		\begin{aligned}
			&\Delta\delta(\Delta t)-\Delta\theta_i(\Delta t)=\lim_{t\rightarrow\Delta t}\int_0^{t}{(\omega_g\Delta\omega(t)-\varpi_i(t))}dt\\
			&=\lim_{\tau\rightarrow\infty}\int_0^\tau{\left(\omega_g\Delta\omega(\tau)-\varpi_i(\tau)\right)}d\tau\\
			&=\lim_{s\rightarrow0}s \mathcal{L}\left\{\Delta\delta(t)-\Delta\theta_i( t)\right\}
			=\frac{K_d}{k_{Ii}U_{ci}}\lim_{s\rightarrow0}s^2\Delta\delta(s)\\
			&=\frac{K_d}{k_{Ii}U_{ci}}\lim_{s\rightarrow0}s^2\frac{\omega_g\Delta\omega(\Delta t)}{s^2}=\frac{K_d}{k_{Ii}U_{ci}}\omega_g\Delta\omega(\Delta t).
		\end{aligned}
	\end{equation}

	 {Therefore,}
	 \begin{equation}
	 	\label{n25}
	 	\begin{aligned}
	 		\hspace{-2.3mm}
	 		\int{(\Delta\delta-\Delta\theta_i)}=\frac{K_d}{k_{Ii}U_{ci}}\omega_g[\Delta\omega(\Delta t_1)+\Delta\omega(\Delta t_2)+...],
	 	\end{aligned}
	 \end{equation}
    {where $\Delta t_1$, $\Delta t_2$, ... represent the individual $\Delta t$ intervals.
    
   Equation \eqref{n25} indicates that the introduction of \(K_d\) effectively reduces \(P_c\) at time \(t_1\) by accumulating the integral term \(\int{(\Delta\delta-\Delta\theta_i)}\). Furthermore, this reduction becomes more pronounced with increasing \(K_d\).
   Consequently, as illustrated in Fig.~\ref{fig_16}, \textbf{the damping effect of classical PLL damping control methods is effectively transferred from the fast time scale to the slow time scale, thereby providing damping for the SynCon's rotor acceleration.}
    	}
	
	\subsection{{Dual Time Scale Stabilization Control Scheme}}
	{Next, we develop a method for designing the damping coefficient $K_d$ to effectively damping SynCon's rotor.}
	
	{The transient energy dissipation of the PLL for the $i$-th converter can be expressed as:}
	\begin{equation}
		\label{n26}
		\begin{aligned}
			\dot V_{ci}=
			-D_{ci}^*{\varpi_i^2}+u_i\varpi_i-K_d\varpi_i^2.
		\end{aligned}
	\end{equation}
	
	{To ensure a negative transient energy variation rate for the PLL ($\dot V_{ci}<0$), $K_d$ is designed to exceed $K_{d1}$, as defined in Eq.\eqref{n27}. When $\varpi_i > 1$, the term $-K_d\varpi_i^2$ can simultaneously compensate for PLL nonlinear damping and suppress non-coherent converter disturbances.}
	\begin{equation}
		\label{n27}
		\begin{aligned}
			K_d>K_{d1}=\frac{k_{Pi}}{k_{Ii}}U_{E}+\max\left\{|u_{i}^{\rm{max}}|,|u_{i}^{\rm{min}}|\right\}.
		\end{aligned}
	\end{equation}
	
	To effectively damp the SynCon's rotor angle increase, \(K_d\) need to be designed to ensure the overdamping condition of the second-order linearized PLL system. This is crucial for fully leveraging its slow-time-scale damping effects. This requirement is expressed as:
	\begin{equation}
		\label{n28}
		K_d>K_{d2}=2\sqrt{k_{Ii}U_{ci}}-k_{Pi}U_{ci}.
	\end{equation}
	
	Furthermore, to ensure that \(P_c\) can decrease to zero during the SynCon's first swing, the integral term \(\int{(\Delta\delta-\Delta\theta_i)}\) must be capable of reaching \(\frac{\pi}{2}\). As indicated by \eqref{n25}:
	\begin{equation}
		\label{n32}
		\int{(\Delta\delta-\Delta\theta_i)}\geq\frac{K_d}{k_{Ii}U_{ci}}\omega_g\Delta\omega_{\max},
	\end{equation}
	where \(\Delta\omega_{\max}\) represents the maximum angular frequency deviation at the PLL convergence instant. \(\Delta\omega_{\max}\) can be estimated as \(\Delta\omega_{\max}\approx\alpha {E_s}\sum_{i=1}^n{I_{ci}}t_f\), where \(t_f\) denotes the fault duration time. Consequently, to meet the condition for \(P_c\) reduction, \(K_d\) must also satisfy:
	\begin{equation}
		\label{n30}
		K_d>K_{d3}=\frac{k_{Ii}U_{ci}}{\omega_g\Delta\omega_{\max}}\frac{\pi}{2}.
	\end{equation}
	
	{Combining conditions \eqref{n27}, \eqref{n28}, and \eqref{n30}, the damping coefficient \(K_d\) is determined as:}
	\begin{equation}
		\label{n31}
		\begin{aligned}
			K_d=\max\left\{K_{d1},K_{d2},K_{d3}\right\}.
		\end{aligned}
	\end{equation}
	
	{To ensure control effectiveness, the PLL damping control is activated during LVRT conditions and remains active for a post-fault duration of several seconds.}

\section{\textcolor{black}{Discussion}}
\textcolor{black}{The preceding analysis is based on assumptions of an infinite receiving-end grid and a single-unit SynCon model. This section investigates how the derived conclusions are affected if these assumptions are relaxed.}

\subsection{\textcolor{black}{Impact of Receiving-End Grid Strength}}
\textcolor{black}{\textbf{1) Impact of the decrease in $Y_g$}}

\textcolor{black}{As previously analyzed, SynCon integration shifts the dominant instability source from PLL to SynCon. This phenomenon is particularly pronounced \textcolor{black}{in weak system scenarios}. Specifically, a reduction in \(Y_g\) directly increases the distribution coefficient \(\alpha\). This leads to an enhanced reallocation of GFLR active current to the SynCon, thereby exacerbating SynCon instability. Conversely, according to \eqref{PMci_with_syncon_change}, while a reduction in \(Y_g\) (an increase in \(L_g\)) could potentially increase the PLL's equivalent mechanical power, the diminishing \((1-\alpha)\) factor effectively mitigates this adverse influence on PLL stability.}

\textcolor{black}{\textbf{2) Impact of the increase in $R_g/L_g$}}

\textcolor{black}{Given that network resistance cannot be neglected, the equivalent mechanical powers for the SynCon and PLL are expressed as:}
\begin{equation}
	\label{PmSynCon}
	P_{c} = \alpha E_s i_c^{d_s }- E_s^2 G_{ss},
\end{equation}
\begin{equation}
	\label{PmPLL}
	\begin{aligned}
		P_{Mci} &= {\omega _g}\left[ {\left( {1 - {\alpha _1}} \right){L_g} + {L_{ci}}} \right]i_{ci}^d
		+ \left[ {\left( {1 - {\alpha}} \right){R_{{g}}} + {R_{ci}}} \right]i_{ci}^q \\
		&\quad + \left( {1 - {\alpha _1}} \right){R_{{g}}}i_{cij}^{{q_i}}
		+ \left( {1 - {\alpha _1}} \right){\omega _g}{L_{{g}}}i_{cij}^{{d_{i}}},
	\end{aligned}
\end{equation}
\textcolor{black}{where $i_{cij}^{{q_i}}$ represents the sum of the $q_i$-axis current components from all other converters; \(G_{ss} \approx R_g Y_g^2\) is the self-conductance term of SynCon node.}

\textcolor{black}{For the SynCon, an increasing \(R_g\) (while maintaining constant \(Y_g\)) leads to increase in \(G_{ss}\) and decrease in \(P_c\), thereby enhancing its stability.
For fast-time scale PLL dynamics, an increasing \(R_g\) amplifies the influence of $q_i$-axis current components (as in \eqref{PmPLL}). However, the \((1-\alpha)\) coefficient  attenuates this impact. Consequently, an increasing \(R_g/X_g\) ratio enhances SynCon stability. The PLL's stability, already improved by SynCon integration, also remains robust. }

\textcolor{black}{Furthermore, as analyzed in Section IV.A, the mechanism by which the additional damping loop in PLLs reduces $P_c$ is not dependent on the system impedance. Therefore, the effectiveness of this method remains robust against variations in grid strength or its \(R_g/X_g\) ratio.}

\subsection{\textcolor{black}{Impact of Multiple GFLR+SynCon clusters}}
\textcolor{black}{\textbf{1) Impact on the fast time scale}}

\textcolor{black}{With multiple SynCons integrated, the equivalent mechanical power of the $i$-th PLL in \eqref{eq3-7b} still applies, but its coupling coefficient $\alpha$ is updated to:}
\begin{equation}
	\label{alpha}
	\begin{aligned}
		\alpha  = \frac{{\sum\nolimits_p {{Y_{sp}}} }}{{{Y_g} + \sum\nolimits_p {{Y_{sp}}} }},
	\end{aligned}
\end{equation}
\textcolor{black}{where $Y_{sp}$ is the admittance between the $p$-th SynCon and PCC.
	Also, the coupling term $E_s \sin(\theta - \delta)$ in $P_{Eci}$ is modified to $\frac{1}{{{Y_g} + \sum\nolimits_p {{Y_{sp}}} }} \sum\nolimits_p {\left[ {{Y_{sp}}{E_{sp}}\sin ({\theta _i} - {\delta _p})} \right]}$, where $E_{sp}$ and $\delta_p$ are the electrical potential and rotor angle of the $p$-th SynCon, respectively.} 
	
	\textcolor{black}{Given the time-scale separation between the rotor dynamics of all SynCons and the PLL still holds, the analysis in Section III.A indicates that $P_{Eci}$ has a relatively small impact on the PLL stability boundary after fault clearance. Instead, the primary influence of multiple SynCons on the PLL is manifested through $P_{Mci}$. As derived from \eqref{alpha}, SynCons integration causes $\alpha$ to increase, which decreases $P_{Mci}$ and enhances PLL stability. This stabilizing effect is further amplified when GFLCs are in closer electrical proximity to any SynCon. In this scenario, the condition $Y_{cp} \gg Y_g$ drives $\alpha$ towards $1$, thereby significantly enhancing PLL stability.}

\textcolor{black}{\textbf{2) Impact on the slow time scale}}

\textcolor{black}{The equivalent mechanical power of the $p$-th SynCon, denoted as $P_{cp}$, is expressed as:}
\begin{equation}
	\label{alphap}
	\begin{aligned}
	P_{cp}={E_{sp}}\sum\limits_{i = 1}^n {{\alpha _{pi}}{I_{ci}}\cos \left( {{\delta _p} - {\varphi _{ci}}} \right)}, 
	\end{aligned}
\end{equation}
\textcolor{black}{where \(\alpha_{pi}\) is the element in the \(p\)-th row and \(i\)-th column of the matrix \( \left( {{{\bf{Y}}_{SC}} - {{\bf{Y}}_{SL}}{{\bf{Y}}_{LL}}^{ - 1}{{\bf{Y}}_{LC}}} \right){{\left( {{{\bf{Y}}_{CC}} - {{\bf{Y}}_{CL}}{{\bf{Y}}_{LL}}^{ - 1}{{\bf{Y}}_{LC}}} \right)}^{ - 1}} \). Here, \(\mathbf{Y}_{SC}\), \(\mathbf{Y}_{SL}\), and \(\mathbf{Y}_{CL}\) are the mutual admittance matrices between synchronous machine nodes and GFLR nodes, synchronous machine nodes and load nodes, and GFLR nodes and load nodes, respectively; while \(\mathbf{Y}_{SS}\), \(\mathbf{Y}_{CC}\), and \(\mathbf{Y}_{LL}\) are the self-admittance matrices for synchronous machine nodes, GFLR nodes, and load nodes, respectively. Physically, $\alpha_{pi}$  is the coefficient that maps the $i$-th GFLC current to the $p$-th SynCon. A shorter electrical distance between SynCon nodes and GFLR nodes results in a larger $\alpha_{pi}$, which, in turn, leads to a larger $P_{cp}$. This determines that a SynCon's dynamics are significantly influenced by its adjacent GFLRs, and its stability degrades with increasing electrical coupling to these GFLRs.
Accordingly, the damping effect of PLLs can still be leveraged to drive $P_{cp}$ to zero after fault clearance, thereby enhancing SynCon stability. This confirms the continued validity of the core conclusions of this study.}

\textcolor{black}{Additionally, the electromagnetic power of the $p$-th SynCon is modified. Specifically, the term ${E_s}{U_g}\alpha\left| {{Y_{g}}} \right|\sin \delta$ is replaced by $\sum\nolimits_j {{B_{pj}}{E_{sp}}{E_{sj}}\sin \left( {{\delta _p} - {\delta _j}} \right)} $, where $B_{pj}$ is the susceptance between the $p$-th and $j$-th SynCons; $E_{sj}$ and $\delta_{j}$ are the voltage and rotor angle of the $j$-th SynCon. This modification reflects the interactive coupling of multiple SynCons' power output. When multiple SynCons are in close electrical proximity, SynCons tend to exhibit coherent behavior under the influence of inter-SynCon power coupling. Conversely, if these clusters are electrically distant, each SynCon is more significantly influenced by its adjacent GFLR's current, potentially leading to incoherent dynamics among SynCons.}

	\section{Controller Hardware in the Loop Experiment}
	\begin{figure}[!t]
		\centering
		\includegraphics[width=2.8in]{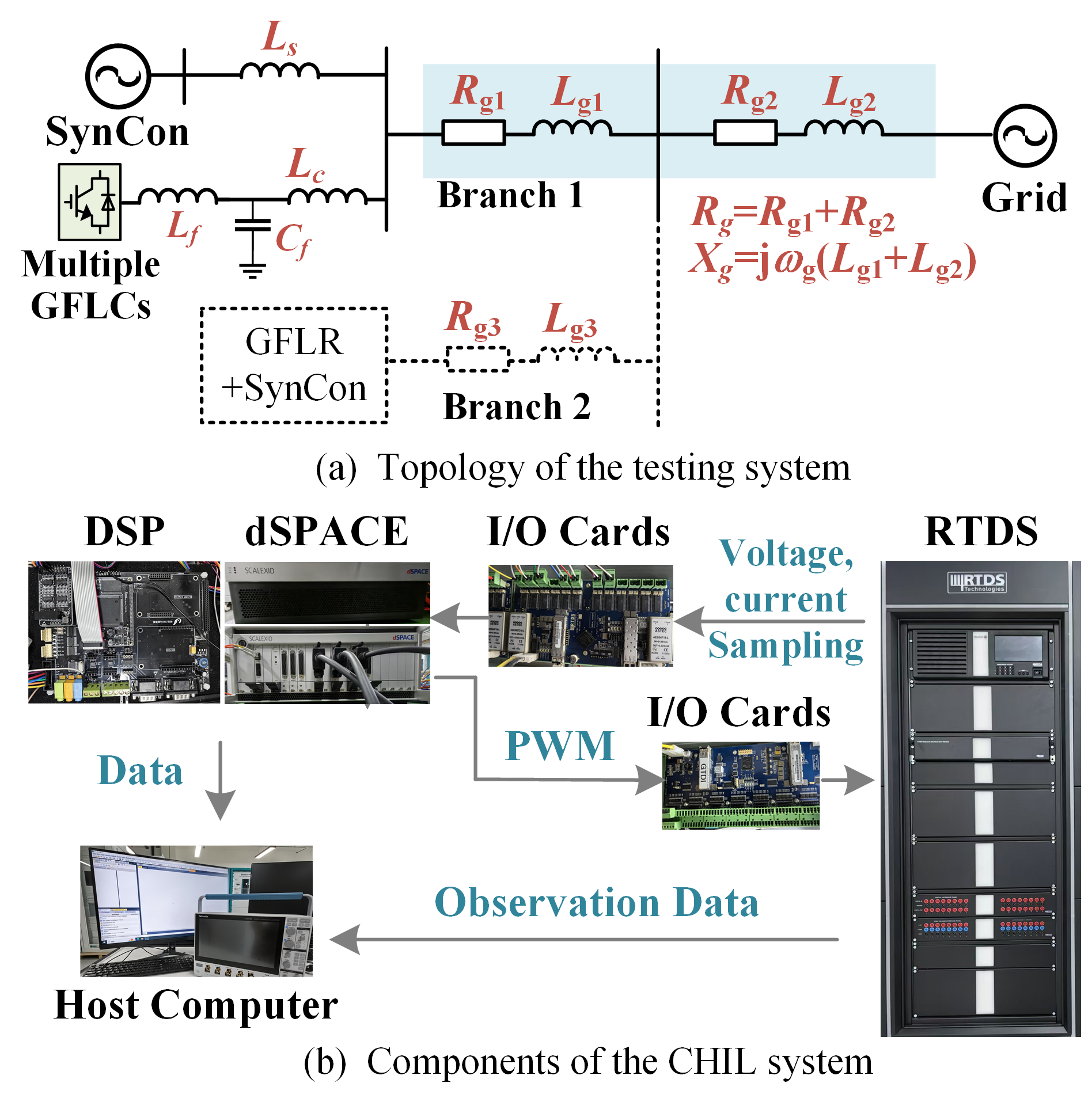}
		\caption{\textcolor{black}{Configuration of the CHIL test system.}}
		\label{fig_11}
	\end{figure}
	To verify the theoretical findings, a controller hardware-in-the-loop platform is constructed, as shown in Fig.~\ref{fig_11}. {The main circuit, comprising GFLCs, SynCon units, transmission lines, and the receiving-end grid, is simulated in RTDS, while the GFLC  controllers are implemented on a DSP-TMS320F28377D and dSPACE board. The three-phase voltage and current measurements at the converter connection points are transmitted to the DSP and Dspace. The PWM signals generated by controllers are then sent back to the RTDS to control the switching of the converter thyristors.} The switching frequency is set to 30kHz.

	\begin{table}[h]
		\centering
		\caption{System Parameters }
		\label{tab:2} 
		\begin{tabular}{lll}
			\toprule
			\textbf{Symbol} & \textbf{Item} & \textbf{Value} \\
			\midrule
			$U_{g}$ & Rated line-to-line grid bus voltage & 100 kV \\
			$S_B$ & Base capacity of the system & 200 MVA\\
			$P_w$ & Active Power of GFLR & 100 MVA\\
			$\omega_g$ & System angular speed & 314 rad/s \\
			$L_f,C_f$ & Inductance, capacitor of filter & 7e$^{-3}$H, 1e$^6\mu$F \\
			$L_{c}$ & Inductance of branches & 0.05 pu \\
			$L_{g1},L_{g2}$ & Inductance of branches & 0.57,0.63 pu \\
			$R_{g1},R_{g2}$ & Resistance of branches & 0.05,0.02 pu \\
			$k_{PC},k_{IC}$ & Proportional,integral gain of current loop & 100,1000 \\
			$k_{P},k_I$ & Proportional,integral gain of PLLs & 12,100 \\ 
			$k_q$ & LVRT reactive current support coefficient & 2 \\
			$S_{sc}$ & Capacity of SynCon & 50MVA \\
			$T_{s}$ & Inertia time constant of SynCon & 6 s \\
			$U_{dc}$ & DC bus voltage of each converter & 5kV \\
			\bottomrule
		\end{tabular}
	\end{table}
	
	\subsection{Dominant Instability Shift with Coherent GFLCs}
To validate the hypothesized shift in the dominant instability source from the PLL to the SynCon as electrical coupling intensifies, we vary the SynCon's grid-connection branch reactance, $L_s$, which directly controls the coupling coefficient $\alpha$. GFLCs are modeled using a multi-machine scaling approach. System parameters are shown in Table~\ref{tab:2}. The fault scenario is a three-phase short circuit on one circuit of the double-circuit line ($0.1\Omega$ fault resistance), cleared after $200$ ms by disconnecting the faulty line.
	
	Figs.~\ref{fig_12} and \ref{fig_13} depict the angle curves and phase trajectories of SynCon and PLLs as $\alpha$ gradually increases from 0 (SynCon disconnected) towards 1. 
	Fig.~\ref{fig_12} reveals that PLLs become unstable immediately after fault clearance at $\alpha=0$ and $\alpha=$0.065. At $\alpha=$0.24, while PLLs do not destabilize immediately post-fault, they lose stability when SynCon's rotor angle reaches 2.05 rad. At this point, PLLs' calculated critical transient energy in \eqref{Lfun} shrinks to just 0.0034, indicating it is on the verge of losing its SEP. In contrast, as $\alpha$ increases further, SynCon's angle stability progressively deteriorates. Eventually, it becomes unstable at $\alpha=$0.64 and 0.77. In these cases, PLLs maintain grid-following behavior, with output angles tracking the angle change caused by SynCon.
	\begin{table}[h]
		\centering
		\caption{\textcolor{black}{Dominant Instability Concern with Different $\alpha$}}
		\label{tab:3}
		\begin{tabular}{cccc}
			\toprule
			\textbf{$\alpha$} & \textbf{\makecell{PLL Stability\\ Metric $\theta_{bi}$}} & \textbf{\makecell{SynCon Stability\\ Metric $P_c$}} & \textbf{\makecell{Instability\\ source}} \\
			\midrule
			0 & 0.501 & 0 & PLLs unstable\\
			0.065 & 0.608 & 3.25 & PLLs unstable\\
			0.24 & 0.828 & 16.8 & \makecell{PLLs unstable\\(loss of SEP))}\\
			0.64 & 1.217 & 54.4 & SynCon unstable\\
			0.77 & 1.332 & 73.15 & SynCon unstable\\
			\bottomrule
		\end{tabular}
	\end{table}
	\begin{figure}[!t]
		\centering
		\includegraphics[width=3.2in]{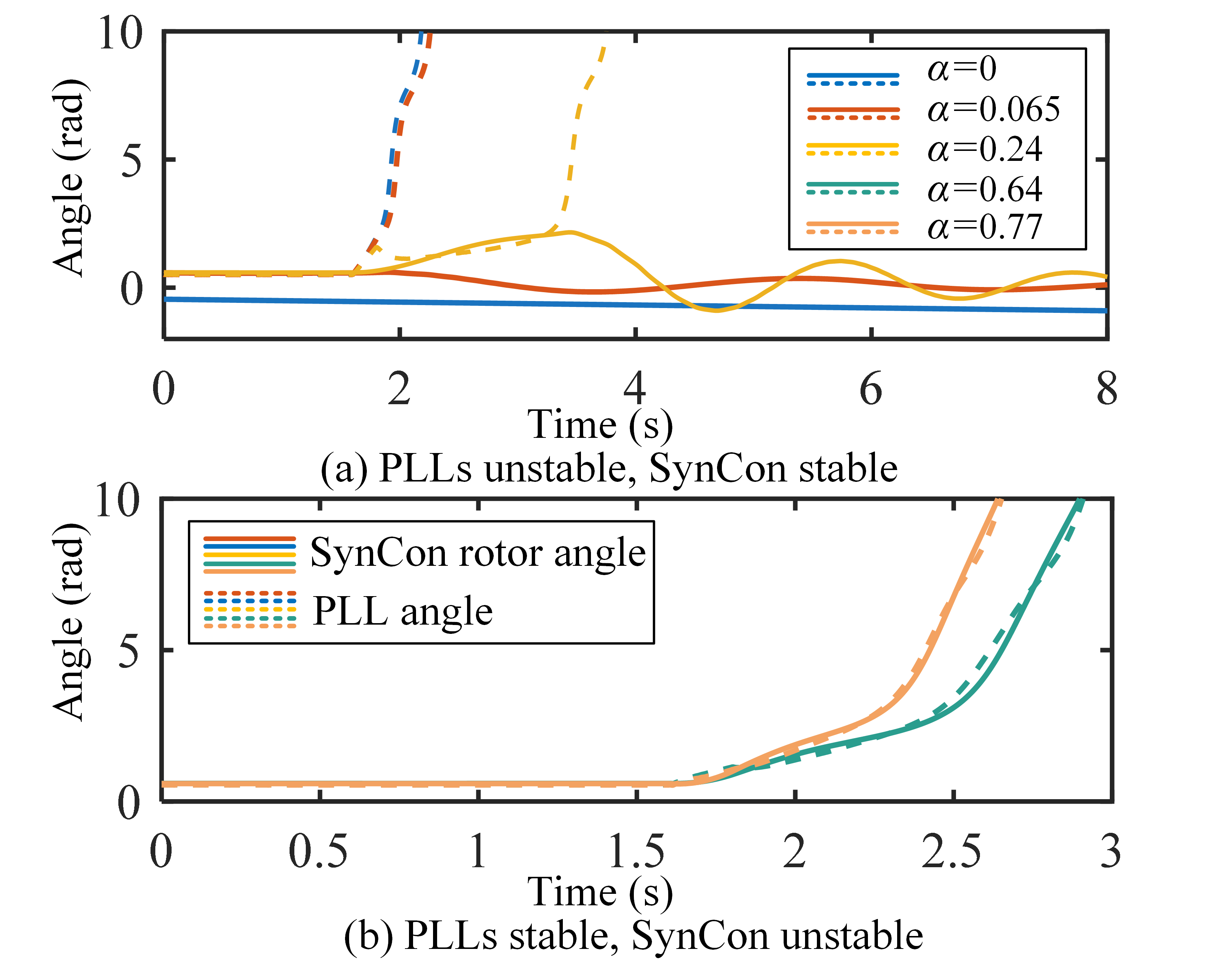}
		\caption{PLLs and SynCon angle curves with different values of $\alpha$.}
		\label{fig_12}
	\end{figure}
	\begin{figure}[!t]
		\centering
		\includegraphics[width=3.1in]{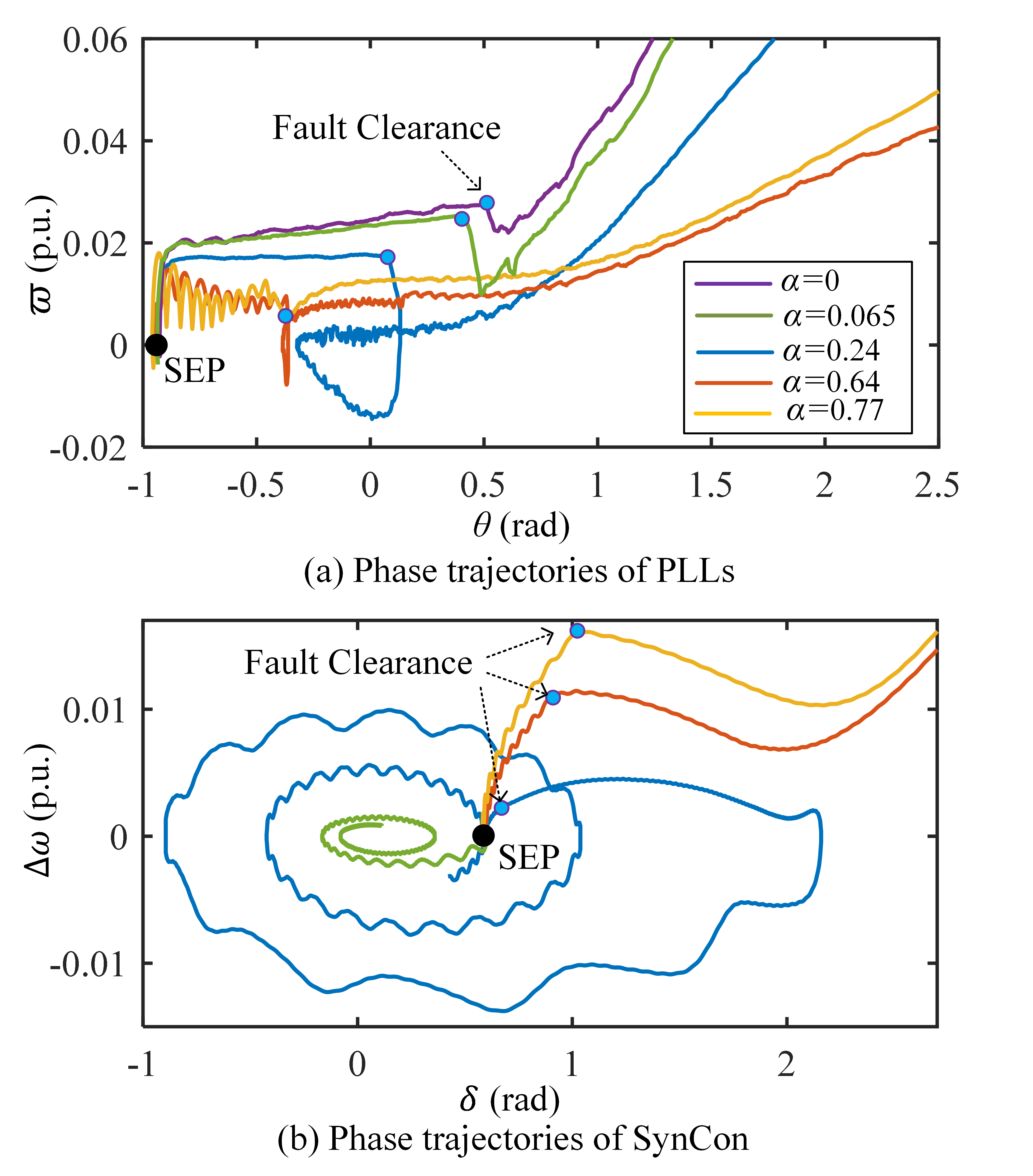}
		\caption{Phase trajectories of SynCon and PLLs.}
		\label{fig_13}
	\end{figure}
	
	Table~\ref{tab:3} presents the calculated stability metric values for PLLs and SynCon at fault clearance. As $\alpha$ increases, the table demonstrates a progressive expansion of PLLs' stability boundary. Concurrently, phase trajectories in Fig.~\ref{fig_13}(a) show a decrease in PLLs' transient energy accumulation during faults, as fault clearance points progressively approaching the SEP. These trends indicate that PLLs' transient stability improves steadily with increasing $\alpha$.
	For SynCon, Table~\ref{tab:3} indicates an increase in its equivalent mechanical power with $\alpha$. This trend, coupled with observations from Fig.~\ref{fig_13}(b), reveals that SynCon's energy accumulation during fault periods intensifies as $\alpha$ increases. This is evidenced by the fault-clearance points being progressively further from the SEP. At $\alpha=$0.065, minimal energy accumulates during faults due to a very small $P_c$. At $\alpha=$0.24, SynCon accumulates more energy during faults, showing tendencies towards instability post-fault clearance. However, the increase in 
	$\delta$ causes the PLL to lose its SEP, which in turn leads to a decrease in $P_c$ and ultimately stabilizes SynCon.
	Consequently, SynCon becomes stable. At higher $\alpha$ values, specifically 0.64 and 0.77, PLLs retain their SEPs despite the increase in $\delta$, and SynCon then exhibits instability.
	
	These observations support the theoretical analysis that the system's dominant instability sources shift from fast to slow time scales as the electrical distance between SynCon and GFLR becomes stronger.

	\subsection{{Dominant Instability Shift and Dual Time Scale Stabilization with Non-coherent GFLCs}}
	{Considering the non-coherent behavior of multiple converters, converters are divided into four clusters, each with its own control parameters. The steady-state output powers for the four clusters are 16MW, 12MW, 32MW, 12MW, with grid-connected branch inductance of 0.09 p.u., 0.16 p.u., 0.02 p.u., 0.03 p.u., respectively. The PLL proportional coefficients are set to 2, 4, 1.5, 10, the integral coefficients to 125, 500, 100, 800, and the LVRT reactive current support coefficients to 0.2, 0.5, 2, and 2.  }
	\begin{figure}[!t]
		\centering
		\includegraphics[width=3.2in]{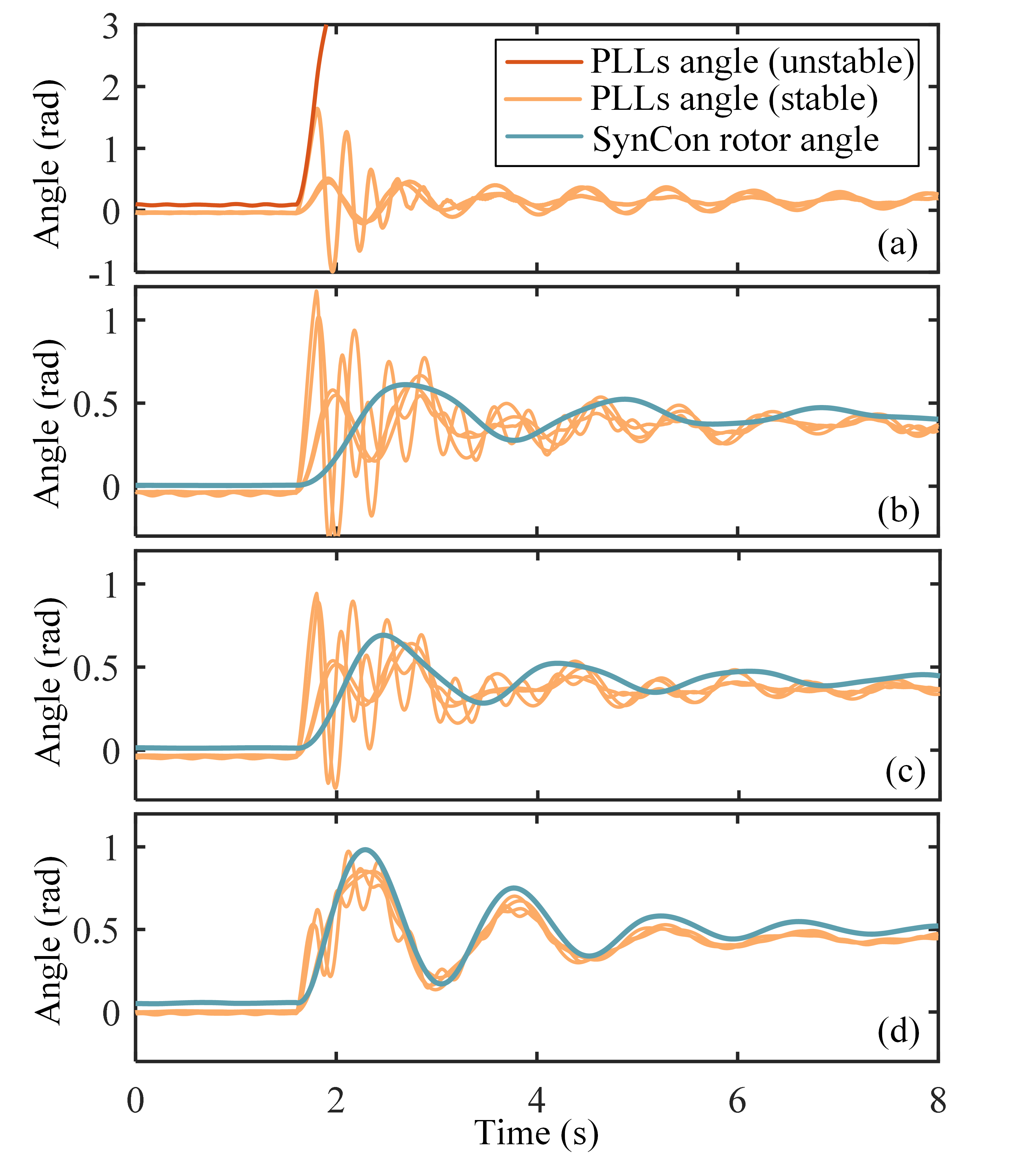}
		\caption{{Variation of angles for 4 converter clusters and SynCon with increasing $\alpha$ at GFLR steady-state output of 72 MW. (a) $\alpha=0$. (b) $\alpha=0.2$. (c) $\alpha=0.3$. (d) $\alpha=0.6$.}}
		\label{fig_14}
	\end{figure}
	\begin{figure}[!t]
		\centering
		\includegraphics[width=3.2in]{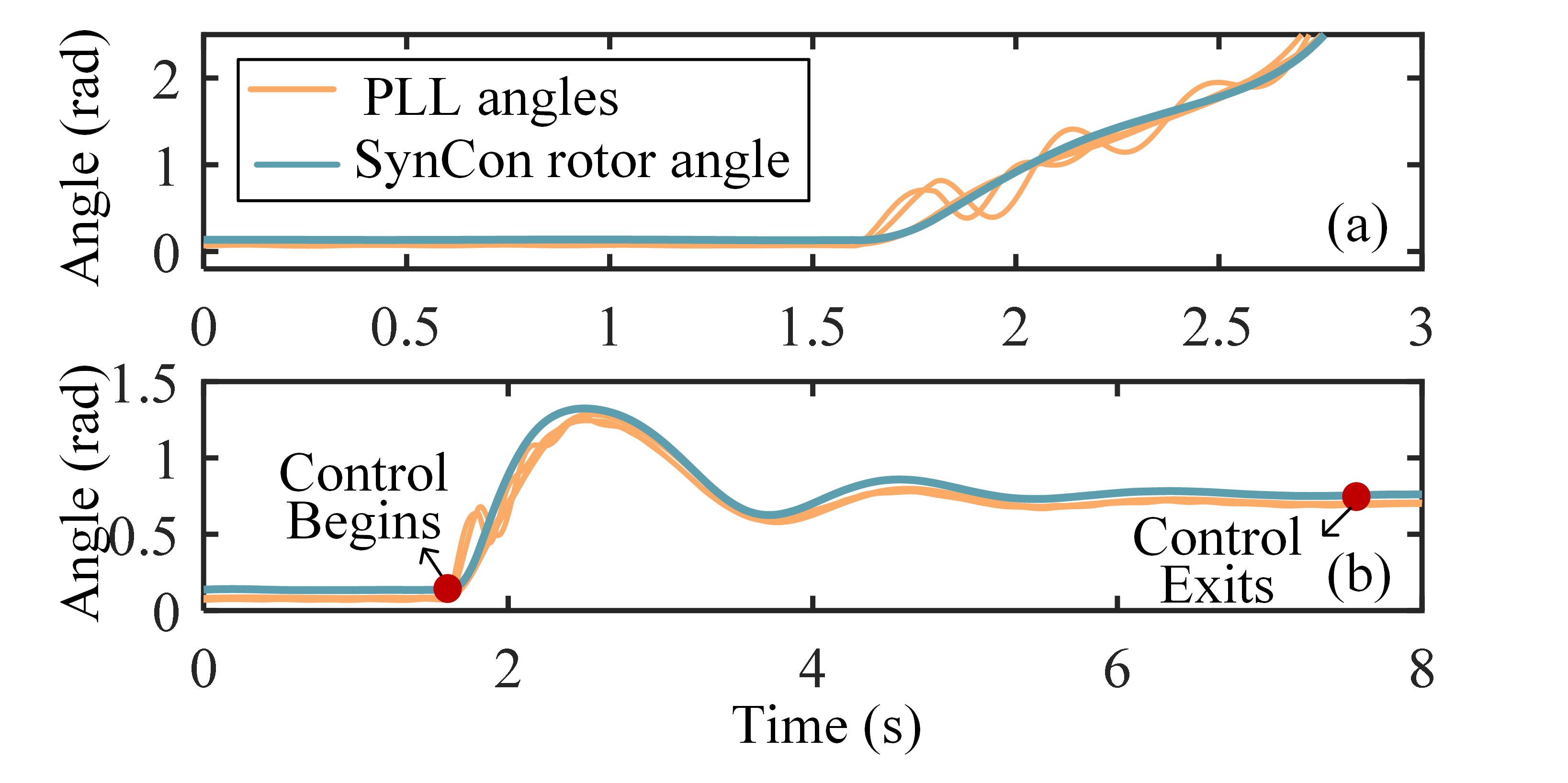}
		\caption{{Comparison of 4 converter clusters and SynCon output angles before and after applying classical PLL damping control at GFLR steady-state output of 100 MW and $\alpha=0.6$. (a) before control. (b) after control.}}
		\label{fig_15}
	\end{figure}
	
	{Fig.~\ref{fig_14} illustrates the impact of increasing $\alpha$ on the angles of the PLLs within the four GFLC clusters and the rotor angle of the SynCon, given a total output of the GFLCs of 72MW. From Fig.~\ref{fig_14} (a), it is observed that without SynCon connection, the four clusters exhibit incoherent behaviors, with one cluster losing stability. However, once the SynCon is connected,  GFLCs' transient stability gradually improves with increasing $\alpha$, and their behavior becomes more coherent. Conversely, the maximum first-swing rotor angle of the SynCon increases as $\alpha$ grows. This indicates that while the SynCon enhances the transient stability of the non-coherent converter clusters, its own stability progressively deteriorates.  }
	
	Fig.~\ref{fig_15} compares the GFLC and SynCon angles before and after applying classical PLL damping control at $\alpha=0.6$, with the total GFLR output reaching 100 MW. The control coefficient $K_{d}=8$ is selected according to the design strategy outlined in Section IV. B. From Fig.~\ref{fig_15}(a), it is observed that the SynCon loses stability after the disturbance, while the PLLs' angles consistently track the SynCon's angle. A comparison of Fig.~\ref{fig_15}(a) and (b) reveals that classical PLL damping control successfully damps the SynCon's acceleration instability. This demonstrates that the GFLR and SynCon power generation combination can achieve internal transient stability without complex measures.
	\subsection{\textcolor{black}{Impact of a Finite Receiving-End Grid}}
	\textcolor{black}{As shown in Fig.~\ref{fig_11}(a), the equivalent resistance and reactance of the receiving-end grid are denoted as $R_{g}$ and $X_{g}$, respectively, with its impedance magnitude given by $Z_{g} =\frac{1}{Y_g}= \sqrt{R_{g}^2 + X_{g}^2}$. This subsection investigates the impact of changes in $Z_{g}$ and the $R_{g}/X_{g}$ ratio on the stability of \textcolor{black}{the co-located system}. The GFLCs are categorized into two groups (GFLC$_1$ and GFLC$_2$), with their respective PLLs denoted as PLL$_1$ and PLL$_2$. The disturbance involves a voltage sag in the receiving-end grid to 0.01 p.u., lasting for 230 ms. The test parameters are listed in Table~\ref{tab:4}.}
	
	\textbf{1) Impact of $Z_{g}$ (with Constant $R_{g}/X_{g}$ Ratio)}
	
	\textcolor{black}{The setup for the test cases is detailed in Table~\ref{tab:5}. Specifically, Case 1 examines the scenario without SynCon at $Z_{g} = 0.66$ p.u. Case 2 considers the system with SynCon integrated at $Z_{g} = 0.66$ p.u. Case 3 then investigates the situation where, with SynCon still integrated, $Z_{g}$ is further increased to $0.75$ p.u.. The test results for these cases are presented in Fig.~\ref{fig_SCR}.}
	
	\begin{table}[h]
		\centering
		\caption{\textcolor{black}{System Parameters}}
		\label{tab:4} 
		\begin{tabular}{lll}
			\toprule
			\textbf{Symbol} & \textbf{Item} & \textbf{Value} \\
			\midrule
			$U_{g}$ & Rated line-to-line grid bus voltage & 220 kV \\
			$S_B$ & Base capacity of the system & 200 MVA\\
			$\omega_g$ & System angular speed & 314 rad/s \\
			$L_f$ & Inductance of filter & 8e$^{-3}$H \\
			$L_{s}$ & Inductance of branches & 0.05 pu \\
			$k_{PC1},k_{IC1}$ &\makecell[l]{Proportional,integral gain of PLL$_1$'s\\ current loop} & 300,500 \\
			$k_{P1},k_{I1}$ & Proportional,integral gain of PLL$_1$ & 1,100 \\ 
			$k_{PC2},k_{IC2}$ & \makecell[l]{Proportional,integral gain of PLL$_2$'s\\ current loop} & 300,1000 \\
			$k_{P2},k_{I2}$ & Proportional,integral gain of PLL$_2$ & 0.3,50 \\ 
			$P_{w1},P_{w2}$ & Active Power of GFLC$_1$,GFLC$_2$ & 100,50 MVA\\
			$S_{sc}$ & Capacity of SynCon & 30MVA \\
			$T_{s}$ & Inertia time constant of SynCon & 4.8s \\
			$U_{dc}$ & DC bus voltage of each converter & 10kV \\
			\bottomrule
		\end{tabular}
	\end{table}
\begin{table}[h]
	\centering
	\caption{\textcolor{black}{Cases Settings}} 
	\label{tab:5}
	\begin{tabular}{ccccc} 
		\toprule
		\textbf{Case} & \textbf{\makecell{$Z_g$\\(p.u.)}} & \textbf{$R_g/X_g$} & \textbf{\makecell{Number of \\SynCon}} & \textbf{Result} \\
		\midrule
		1 & 0.66& 0.005 & 0 & \makecell{PLLs  unstable} \\ 
		2 & 0.66& 0.005 & 1 & PLLs and SynCon stable \\
		3 & 0.75& 0.005 & 1 & PLLs stable, SynCon unstable \\
		4 & 0.75 & 0.1 & 0 & PLL$_1$ unstable, PLL$_2$ stable \\
		5 & 0.75& 0.1 & 1 & PLLs stable, SynCon unstable  \\
		6 & 0.75& 0.5 & 1 & PLLs and SynCon stable \\
		\bottomrule
	\end{tabular}
\end{table}
	\textcolor{black}{From Fig.~\ref{fig_SCR}(a), it is evident that without SynCon and at $Z_{g} = 0.66$ p.u., both PLL$_1$ and PLL$_2$ exhibit instability. After SynCon integration, with $\alpha$ becomes $0.93$, Fig.~\ref{fig_SCR}(b) demonstrates that both the SynCon and PLLs maintain stability post-disturbance. However, upon increasing $Z_{g}$, the SynCon becomes unstable. Notably, in this scenario, the PLLs do not become unstable independently but rather track the unstable SynCon. This crucial distinction is drawn by comparing the timing and rate of development of PLL instability in Fig.~\ref{fig_SCR}(a) and (c). Overall, SynCon integration enhances PLL stability, shifts the system's dominant instability source from the PLLs to the SynCon. Moreover, it reveals that an increase in $Z_{g}$
	has a more pronounced impact on SynCon stability.}
		
	\begin{figure}[!t]
		\centering
		\includegraphics[width=3.2in]{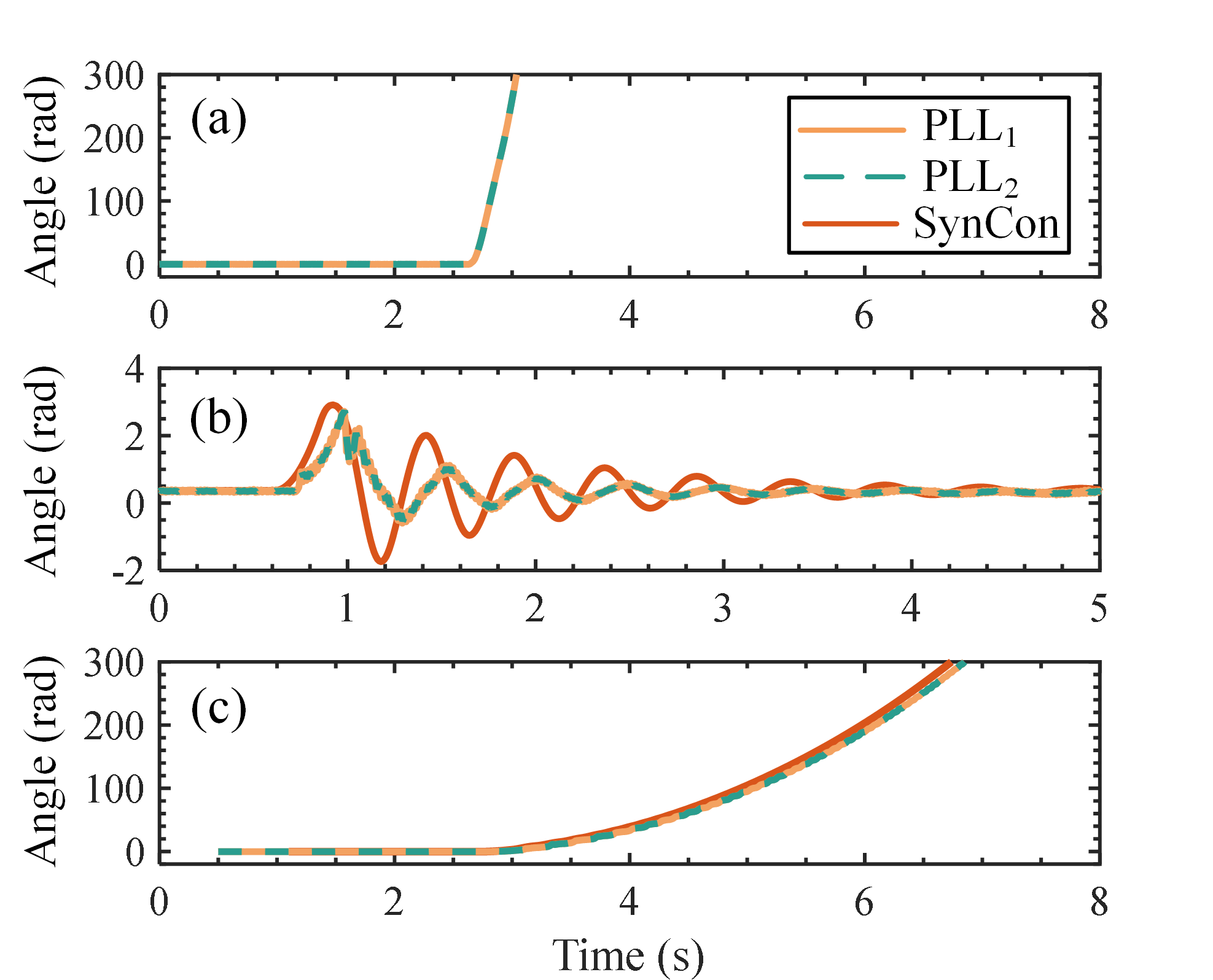}
		\caption{\textcolor{black}{{Test results for Cases 1, 2, and 3. (a) PLL angles in Case 1; (b) PLL and SynCon angles in Case 2; (c) PLL and SynCon angles in Case 3.}}}
		\label{fig_SCR}
	\end{figure}
		\begin{figure}[!t]
		\centering
		\includegraphics[width=3.3in]{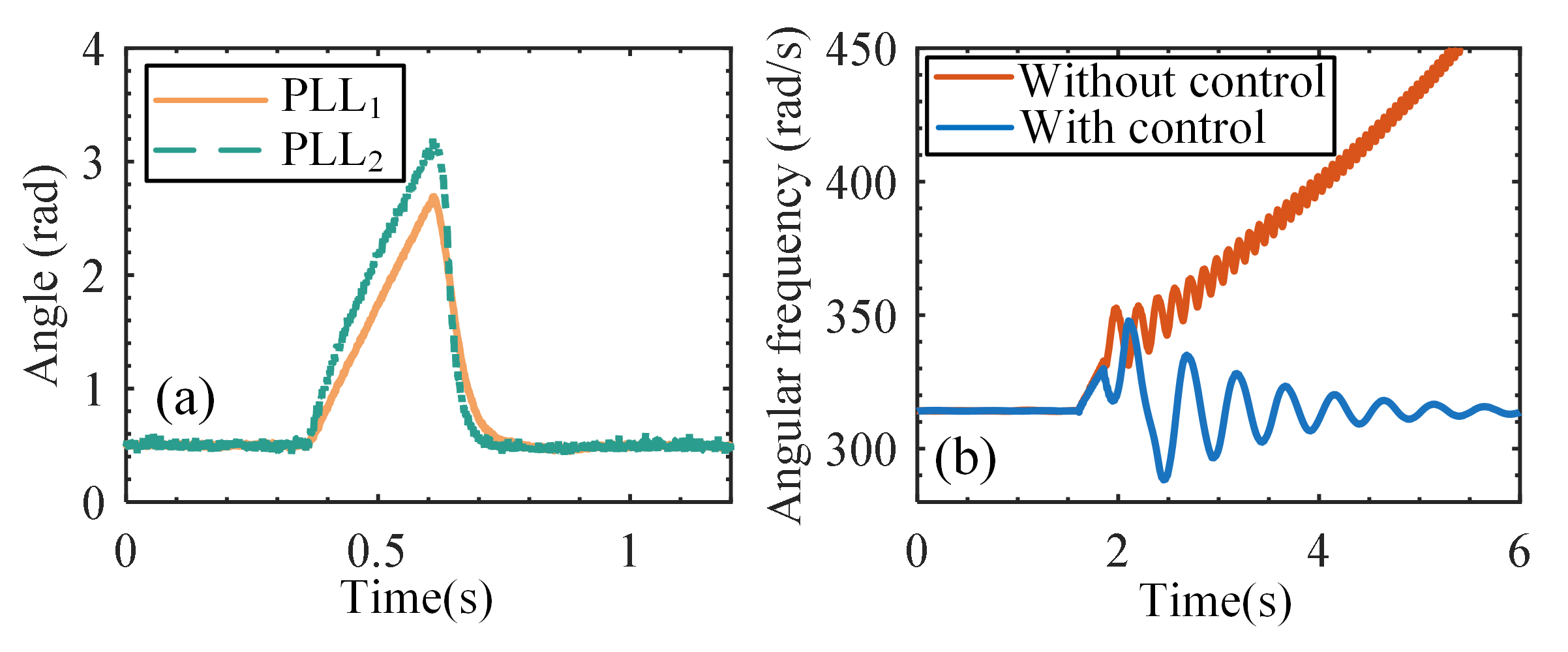}
		\caption{\textcolor{black}{{Experimental results for Cases 1 and 3 after applying the PLL damping loop. (a) PLL angles in Case 1 with damping control; (b) SynCon rotor angular frequency in Case 3 with and without damping control.}}}
		\label{fig_SCRcontrol}
	\end{figure}
	\textcolor{black}{Fig.~\ref{fig_13} presents the post-disturbance test results for the system when both PLL$_1$ and PLL$_2$ are equipped with the damping loop illustrated in Fig.~\ref{fig_10}. The damping coefficients $K_d$ for PLL$_1$ and PLL$_2$ are set to $43$ and $31$, respectively. Specifically, Fig.~\ref{fig_13}(a) displays the PLLs' angles in Case 1 after control implementation, while Fig.~\ref{fig_13}(b) compares the SynCon rotor angular frequency before and after control implementation in Case 3. Evidently, this enhanced damping control for the PLLs ensures their stability before SynCon integration. Furthermore, upon SynCon integration, its stabilizing effect extends to the SynCon, effectively suppressing its rotor angle instability.}
	
	\textbf{2) Impact of $R_{g}/X_{g}$ ($Z_{g}$ constant)}

	\textcolor{black}{To investigate the influence of $R_{g}/X_{g}$, the cases are configured as detailed in Table~\ref{tab:5}. Specifically, Cases 4, 5, and 6 correspond to scenarios with $R_{g}/X_{g}=0.1$ (without SynCon), $R_{g}/X_{g}=0.1$ (with SynCon integration), and $R_{g}/X_{g}=0.5$ (with SynCon), respectively. The test results are presented in Fig.~\ref{fig_RX}(a), (b), and (c).}
	
	\textcolor{black}{From Fig.~\ref{fig_RX}(a), it is observed that before SynCon integration, at $R_{g}/X_{g}=0.1$, PLL$_1$ becomes unstable after the disturbance, while PLL$_2$ remains stable. Subsequently, with SynCon integrated (Case 5), the SynCon rotor angle exhibits instability, as shown in Fig.~\ref{fig_RX}(b). A comparison of the PLL dynamic time scales in Fig.~\ref{fig_RX}(a) and Fig.~\ref{fig_RX}(c) indicates that, for Case 5, both PLL$_1$ and PLL$_2$ remain stable and track the unstable SynCon's angle. When $R_{g}/X_{g}$ is increased to 0.5 (Case 6), both PLLs and the SynCon remain stable, as depicted in Fig.~\ref{fig_RX}(c).
	It can thus be concluded that when $R_{g}/X_{g} \neq 0$, SynCon integration consistently shifts the system's dominant instability source from fast-time-scale PLLs to slow-time-scale SynCon. Furthermore, an increase in $R_{g}/X_{g}$ generally enhances system stability.}
	
	\textcolor{black}{Figs~\ref{fig_RXcontrol}(a) and (b) demonstrate the effectiveness of applying PLL damping control in Case 4 and Case 5, respectively. These figures show that the PLL damping control can still effectively stabilize the PLLs and, after SynCon integration, also achieve stabilization of the SynCon.}

	\begin{figure}[!t]
		\centering
		\includegraphics[width=3.2in]{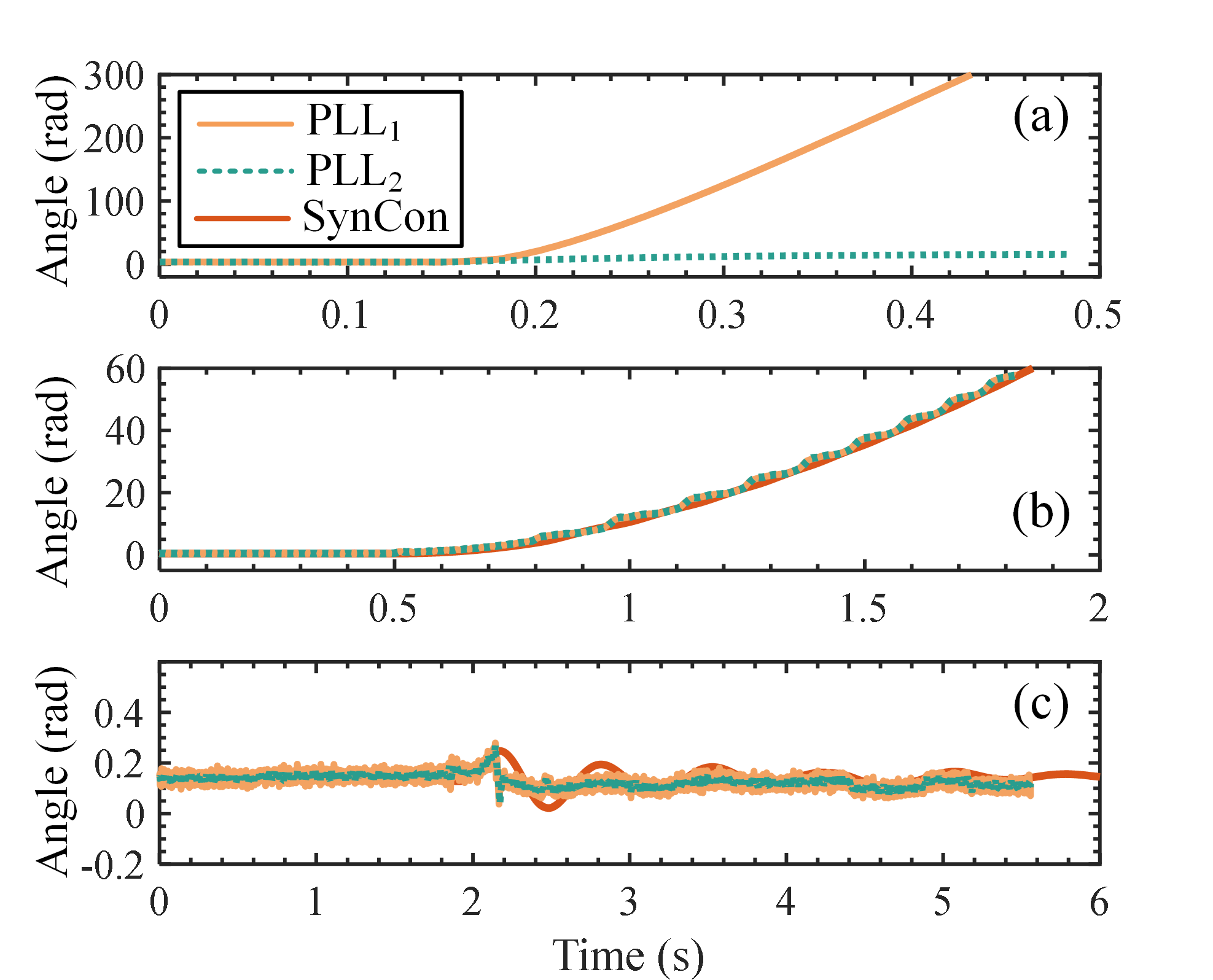}
		\caption{\textcolor{black}{{Experimental results for Cases 4, 5, and 6. (a) PLL angles in Case 4; (b) PLL and SynCon angles in Case 5; (c) PLL and SynCon angles in Case 6.}}}
		\label{fig_RX}
	\end{figure}
	\begin{figure}[!t]
		\centering
		\includegraphics[width=3.4in]{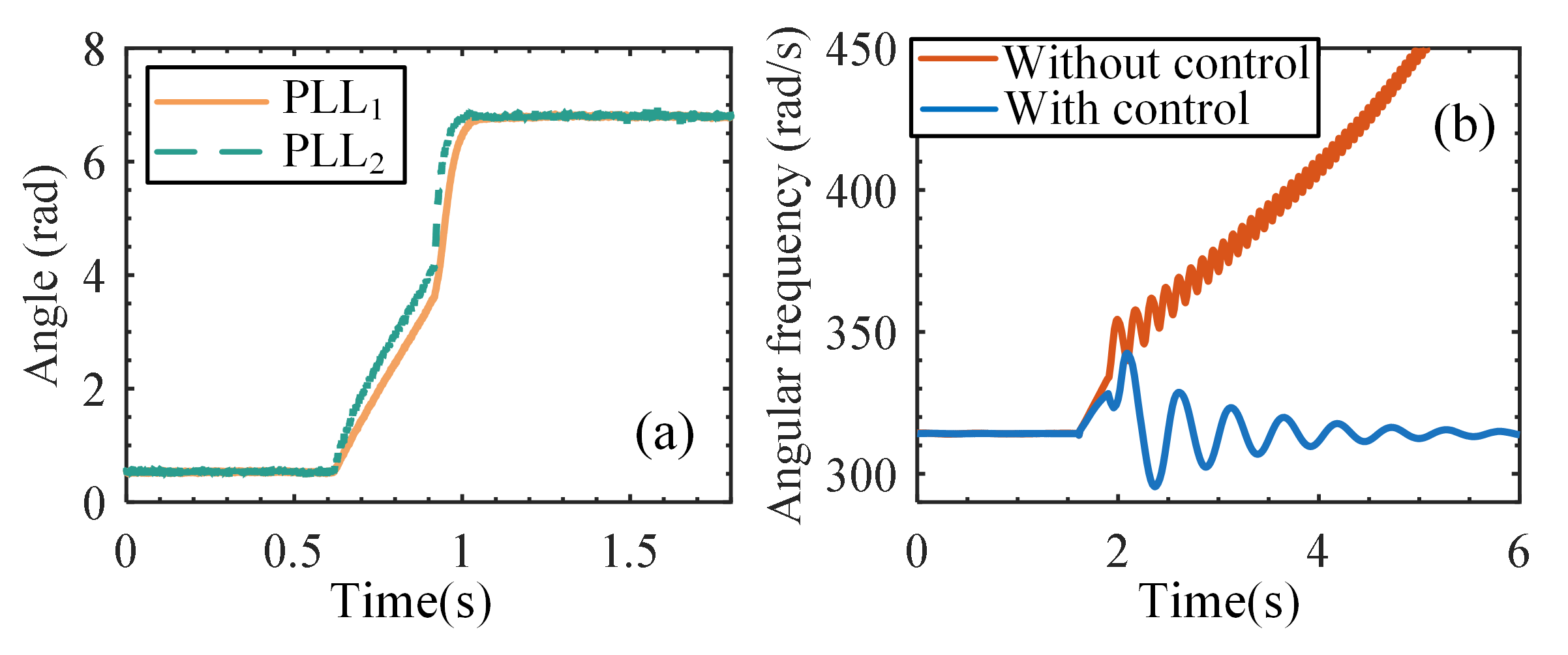}
		\caption{\textcolor{black}{{Experimental results for Cases 4 and 5 after applying the PLL damping loop. (a) PLL angles in Case 4 with damping control; (b) SynCon angular frequency in Case 5 with and without damping control.}}}
		\label{fig_RXcontrol}
	\end{figure}

	\subsection{\textcolor{black}{Impact of Multiple Parallel GFLR+SynCons Clusters}}	
	\textcolor{black}{To investigate the influence of multiple parallel SynCons on the conclusions drawn in this paper, branch 2 is integrated into the test system, as shown in Fig.~\ref{fig_11}(b). \textcolor{black}{Consequently, the system comprises two parallel SynCon+GFLR clusters.} The SynCon and GFLC in the first cluster are denoted as SynCon$_1$ and GFLC$_1$, respectively; similarly, SynCon$_2$ and GFLC$_2$ are designated for the second cluster. The PLLs for GFLC$_1$ and GFLC$_2$ are PLL$_1$ and PLL$_2$, respectively. The parameters for both GFLCs are consistent with those in Table~\ref{tab:5}. SynCon$_1$ and SynCon$_2$ have capacities of 30 MVA and 50 MVA, respectively, with inertia time constants of 6s and 4.8s. Fault settings are identical to those described in Section VI. C.}
	
	\textcolor{black}{Cases 7 and 9 in Table~\ref{tab:6} are configured based on varying electrical distances between the clusters without SynCon integration. Cases 8 and 10 represent the scenarios where SynCons are integrated into Case 7 and Case 9, respectively. Fig.~\ref{fig_multiSynCOn}(a) and (b) present the results for Case 7 and 8, respectively, while Fig.~\ref{fig_multiSynCOn}(c) shows the angular velocities of the SynCons after applying PLL damping control based on Case 8. Fig.~\ref{fig_multiSynCOn}(d), (e), and (f) display the corresponding results for Case 9, Case 10, and Case 10 with PLL damping control, respectively.}
	
	\begin{table}[h]
		\centering
		\caption{\textcolor{black}{Cases Settings}}
		\label{tab:6} 
		
		\begin{tabular}{ 
				>{\centering\arraybackslash}m{0.4cm} 
				>{\centering\arraybackslash}m{0.6cm} 
				>{\centering\arraybackslash}m{0.6cm} 
				>{\centering\arraybackslash}m{0.6cm} 
				>{\centering\arraybackslash}m{0.6cm} 
				>{\raggedright\arraybackslash}m{3.5cm} 
			} 
			\toprule
			\textbf{\makecell{Case}} & 
			\textbf{\makecell{$L_{g1}$\\(p.u.)}} & 
			\textbf{\makecell{$L_{g2}$\\(p.u.)}} & 
			\textbf{\makecell{$L_{g3}$\\(p.u.)}} & 
			\textbf{\makecell{SynCon\\Number}} & 
			\multicolumn{1}{c}{\textbf{\makecell{Result}}} \\ 
			\midrule
			
			7 & 0.27 & 0.56 & 0.51 & 0 & PLL$_1$ unstable, PLL$_2$ stable \\
			8 & 0.27 & 0.56 & 0.51 & 2 & PLLs and SynCons unstable \\
			9 & 0.66 & 0.43 & 0.39 & 0 & PLL$_1$ unstable, PLL$_2$ stable \\
			10 & 0.66 & 0.43 & 0.39 & 2 & SynCon$_1$ and PLL$_1$ unstable, SynCon$_2$ and PLL$_2$ stable \\ 
			
			\bottomrule
		\end{tabular}
	\end{table}

    \textcolor{black}{For Cases 7 and 8, Fig.~\ref{fig_multiSynCOn}(a) illustrates that before SynCon integration, PLL$_1$ becomes unstable, whereas PLL$_2$ remains stable after the disturbance. Upon SynCon integration, as depicted in Fig.~\ref{fig_multiSynCOn}(b), both SynCon$_1$ and SynCon$_2$ exhibit instability, with their respective PLLs (PLL$_1$ and PLL$_2$) closely tracking the unstable SynCons. Then, after the implementation of the PLL damping loop, both SynCon$_1$ and SynCon$_2$ recover stability post-disturbance, as illustrated in Fig.~\ref{fig_multiSynCOn}(c).}
    
    \textcolor{black}{For Cases 9 and 10, as shown in Fig.~\ref{fig_multiSynCOn}(d), before SynCon integration, PLL$_1$ becomes unstable, while PLL$_2$ remains stable after the disturbance. With SynCons integrated, SynCon$_1$ becomes unstable, and its adjacent PLL$_1$ tracks this instability. Conversely, SynCon$_2$ remains stable, with its adjacent PLL$_2$ also remaining stable and tracking it, as depicted in Fig.~\ref{fig_multiSynCOn}(e). Subsequently, after implementing the PLL damping loop, both SynCon$_1$ and SynCon$_2$ recover stability post-disturbance, as illustrated in Fig.~\ref{fig_multiSynCOn}(f).}
    
    \textcolor{black}{Therefore, in scenarios involving multiple GFLR+SynCon clusters connected in parallel at different electrical distances, the integration of SynCons also shifts the system's dominant instability source from the PLLs to the SynCons. This aligns with the conclusions drawn in this paper. It is also observed that SynCons at different electrical distances may either maintain synchronism or exhibit varied stability (some unstable, some stable). Concurrently, the application of additional damping control to the PLLs is also proven effective in suppressing rotor angle instability in multiple SynCons.}
	
	\begin{figure}[!t]
		\centering
		\includegraphics[width=3.3in]{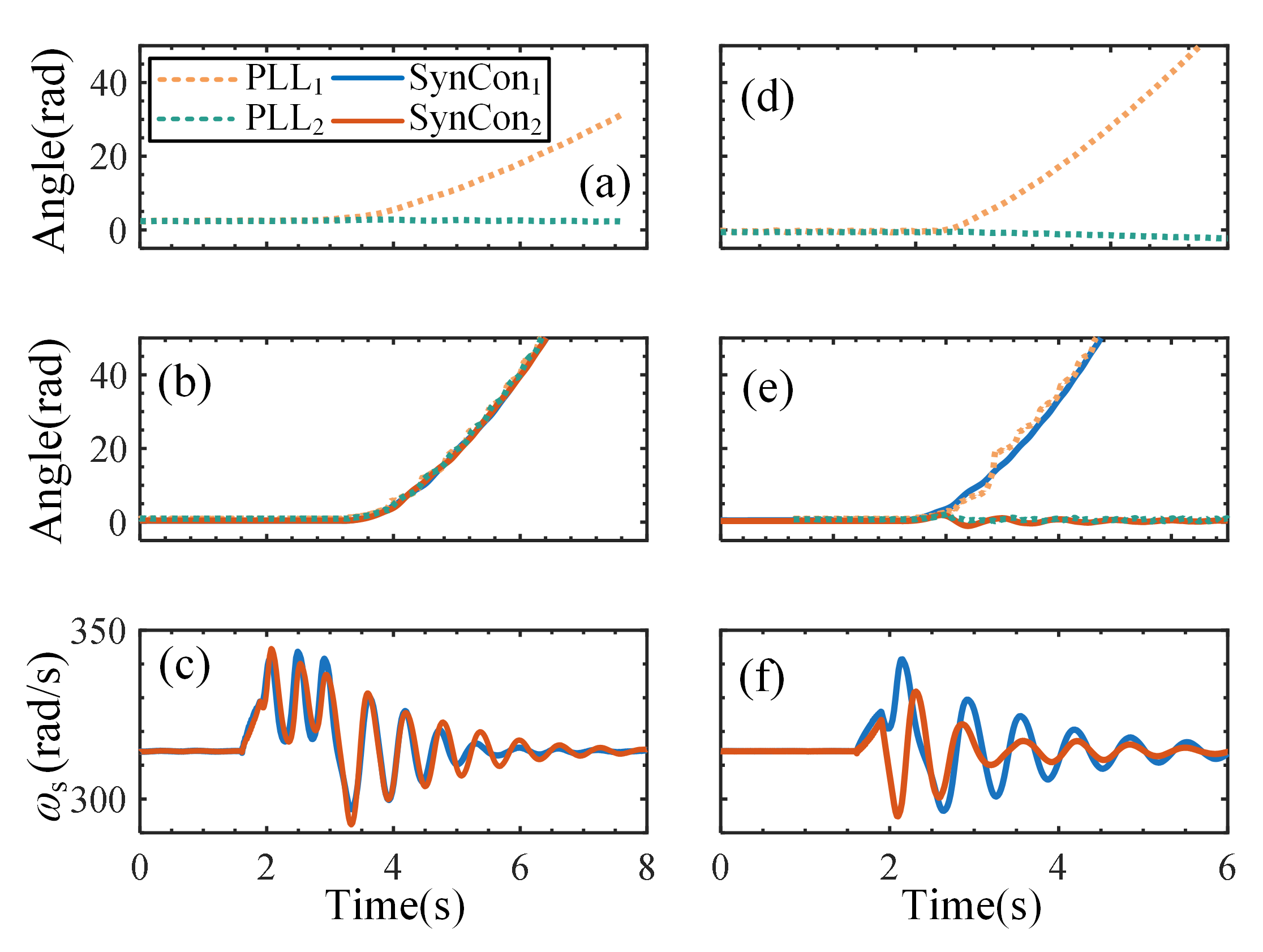}
		\caption{\textcolor{black}{Experimental results for Cases 7, 8, 9, and 10. (a) PLL angles in Case 7; (b) PLL and SynCon angles in Case 8; (c) SynCons' angular frequency in Case 7 with damping control; (d) PLL angles in Case 9; (e) PLL and SynCon angles in Case 10; (f) SynCons' angular frequency in Case 10 with damping control.}}
		\label{fig_multiSynCOn}
	\end{figure}

	\section{Conclusion}
	\textcolor{black}{This study investigates the dual-time-scale transient stability of co-located GFLR–SynCon systems, focusing on the mechanism of the primary transient instability source's shift after SynCon integration and proposing a simple yet practical stabilization approach. The main findings are:}
	\begin{enumerate}
			\item \textcolor{black}{The voltage-source characteristics of the SynCon and its time-scale separation from PLL dynamics can provide a voltage reference for the PLL during faults, enlarge the PLL stability boundary, and mitigate adverse coupling effects among multiple converters. However, the dominant instability concern shifts from the PLLs to the SynCon’s rotor, as their electrical coupling tightens.}
			\item \textcolor{black}{A well-tuned classical PLL damping control is sufficient to stabilize the SynCon. This work reveals that PLLs' damping effect at the fast time scale can be transferred to the slow time scale, thereby damping the rotor acceleration of the SynCon.}
	\end{enumerate}

	These findings underscore the promise of GFLR–SynCon integration as a viable strategy for enhancing transient synchronization stability in modern power systems. \textcolor{black}{By characterizing the fundamental interaction between a current source and a voltage source, this study offers insights applicable to a broader class of heterogeneous systems, such as co-located GFLR and unsaturated GFM systems.}

	\begin{IEEEbiography}[{\includegraphics[width=1in,height=1.2in,clip,keepaspectratio]{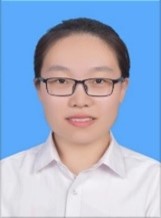}}]{Bingfang Li}(S'23--M'26) received the B.S. degree in electrical engineering from North China Electric Power University, Baoding, China, in 2022, and is currently working toward the Ph.D. degree with Xi'an Jiaotong University. Her main fields of interest include Power system stability analysis and control.
	\vspace{-20pt}
	\end{IEEEbiography}
	\vspace{-30pt}
	\begin{IEEEbiography}[{\includegraphics[width=1in,height=1.2in,clip,keepaspectratio]{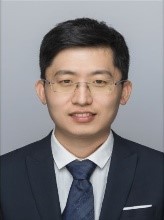}}]{Songhao Yang}(S'18--M'19--SM'24) was born in Shandong, China, in 1989. He received the B.S. and Ph.D. degrees in electrical engineering from Xi’an Jiaotong University, Xi’an, China, in 2012 and 2019, respectively, and the Ph.D. degree in electrical and electronic engineering from Tokushima University, Tokushima, Japan, in 2019. He is currently an Associate Professor with Xi’an Jiaotong University. His research focuses on power system stability analysis and control.
	\vspace{-20pt}
	\end{IEEEbiography}
	\vspace{-30pt}
	\begin{IEEEbiography}[{\includegraphics[width=1in,height=1.2in,clip,keepaspectratio]{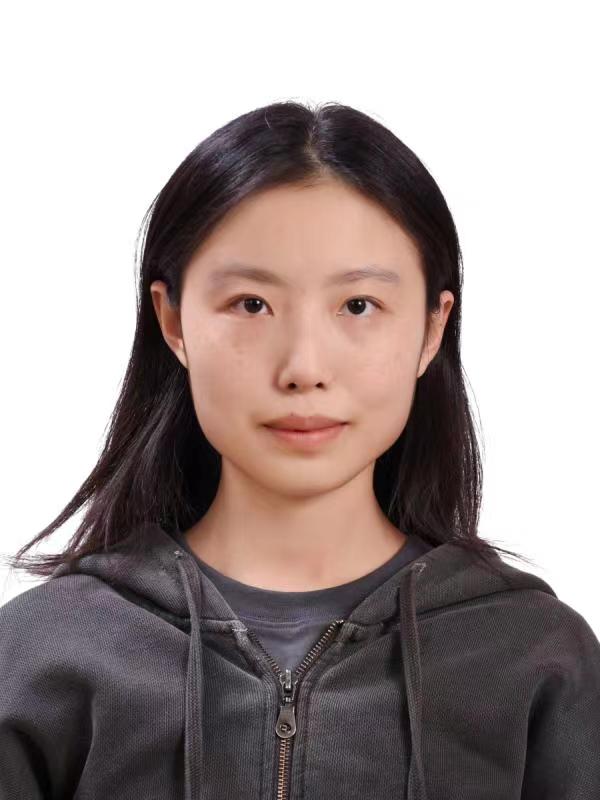}}]{Qinglan Wang} received the B.S. degree in electrical engineering from Xi’an Jiaotong University, Xi’an, China, in 2024, and is currently working toward the M.S. degree with Xi’an Jiaotong University. Her main fields of interest include Power system stability analysis and control.
	\vspace{-20pt}
	\end{IEEEbiography}
	\vspace{-30pt}
	\begin{IEEEbiography}[{\includegraphics[width=1in,height=1.2in,clip,keepaspectratio]{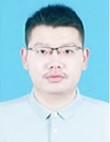}}]{Xu Zhang} (S'21), received the B.S. degree from Xi’an Jiaotong University, Xi’an, China, in 2021, and is currently working toward the Ph.D. degree with Xi’an Jiaotong University. Her main fields of interest include Power system voltage stability analysis.
	\vspace{-20pt}
	\end{IEEEbiography}
	\vspace{-30pt}
	\begin{IEEEbiography}[{\includegraphics[width=1in,height=1.2in,clip,keepaspectratio]{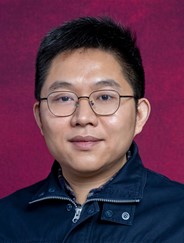}}]{Huan Xie} is working as an electrical engineer at State Grid Jibei Electric Power Research Institute, Beijing, China. He received the B.Sc. and M.Sc. in Electrical Engineering and Automation from Hohai University in 2001 and 2014, and received Ph.D. degree in Electrical Engineering and Automation from Xi'an Jiao Tong University, Xi'an, China in 2008. His areas of interest include power system stability and control.
	\vspace{-20pt}
	\end{IEEEbiography}
	\vspace{-30pt}
	\begin{IEEEbiography}[{\includegraphics[width=1in,height=1.2in,clip,keepaspectratio]{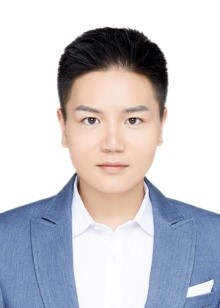}}]{Chuan Qin} is working as an electrical engineer at State Grid Jibei Electric Power Research Institute, Beijing, China. He received the B.Sc. in Electrical Engineering and Automation from University of Electronic Science and Technology of China, Chengdu, China in 2018, and received M.Sc. in Electrical Engineering and Automation from Xi'an Jiaotong University, Xi'an, China in 2021. His areas of interest include power system stability and control.
	\vspace{-20pt}
	\end{IEEEbiography}
	\vspace{-30pt}
	\begin{IEEEbiography}[{\includegraphics[width=0.96in,height=1.2in,clip,keepaspectratio]{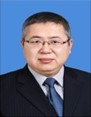}}]{Zhiguo Hao} (M'10-SM'23), was born in Ordos, China, in 1976. He received the B.Sc. and Ph.D. degrees in electrical engineering from Xi’an Jiaotong University, Xi’an, China, in 1998 and 2007, respectively. He is currently a Professor with the Electrical Engineering Department, Xi’an Jiaotong University. His research focuses on power system protection and control.
	\end{IEEEbiography}

\begin{thebibliography}{99}
		\bibliographystyle{IEEEtran}
		\bibitem{1} \textcolor{black}{Y. Xiong, H. Wu, Y. Li, and X. Wang, “Comparison of Power Swing Characteristics and Efficacy Analysis of Impedance-Based Detections in Synchronous Generators and Grid-Following Systems,” {\it{IEEE Trans. Power Syst.}}, vol. 40, no. 3, pp. 2545–2556, May 2025, doi: 10.1109/TPWRS.2024.3469235.}
		\bibitem{2} K. Strunz, K. Almunem, C. Wulkow, M. Kuschke, M. Valescudero and X. Guillaud, "Enabling 100\% Renewable Power Systems Through Power Electronic Grid-Forming Converter and Control: System Integration for Security, Stability, and Application to Europe," {\it{Proc. IEEE}} , vol. 111, no. 7, pp. 891-915, July 2023, doi: 10.1109/JPROC.2022.3193374.
		
		
		\bibitem{4} \textcolor{black}{L. Bao, L. Fan, and Z. Miao, “Maximizing Synchronous Condensers’ Capability to Stabilize Inverter-Based-Resource-Penetrated Grids,” {\it{IEEE Trans. Energy Convers.}}, vol. 40, no. 1, pp. 93–105, Mar. 2025, doi: 10.1109/TEC.2024.3422132.}
		\bibitem{5} \textcolor{black}{Florence School of Regulation. (2022, July). “Blackout hits Spain and Portugal: what happened and what’s next.” [Online]. Available: https://fsr.eui.eu/blackout-hits-spain-and-portugal-what-happened-and-whats-next/}
		\bibitem{6} \textcolor{black}{Q. Qu, X. Xiang, K. Xin, Y. Liu, W. Li, and X. He, “Transient Stability Analysis of Hybrid GFL-GFM System Considering Various Damping Effects,” {\it{IEEE Trans. Ind. Electron.}}, early access, doi: 10.1109/TIE.2025.3581260.}
	
		\bibitem{8} \textcolor{black}{H. Xin, C. Liu, X. Chen, Y. Wang, E. Prieto-Araujo, and L. Huang, ``How Many Grid-Forming Converters Do We Need? A Perspective From Small Signal Stability and Power Grid Strength," {\it{IEEE Trans. Power Syst.}}, vol. 40, no. 1, pp. 623–635, Jan. 2025, doi: 10.1109/TPWRS.2024.3393877.}
		\bibitem{9} \textcolor{black}{C. Luo et al., “Two-Stage Transient Control for VSG Considering Fault Current Limitation and Transient Angle Stability,” {\it{IEEE Trans. Ind. Electron.}}, vol. 71, no. 7, pp. 7169–7179, Jul. 2024, doi: 10.1109/TIE.2023.3292877.}
	
		\bibitem{11} R. W. Kenyon, A. Hoke, J. Tan and B. -M. Hodge, "Grid-Following Inverters and Synchronous Condensers: A Grid-Forming Pair?," in {\it{2020 PSC}}, Clemson, SC, USA, 2020, pp. 1-7. 
		
		
		\bibitem{14} \textcolor{black}{S. Ghimire, K. Vatta Kkuni, E. D. Guest, K. H. Jensen, and G. Yang, “Impact of Synchronous Condensers on Small-Signal Stability of Offshore Wind Power Plants,” {\it{IEEE Access}}, vol. 12, 2024, Art. no. 168018, doi: 10.1109/ACCESS.2024.3497669.}
		\bibitem{15} S. Hadavi, J. Saunderson, A. Mehrizi-Sani, and B. Bahrani, “A Planning Method for Synchronous Condensers in Weak Grids Using Semi-Definite Optimization,” {\it{IEEE Trans. Power Syst.}}, vol. 38, no. 2, pp. 1632–1641, Mar. 2023, doi: 10.1109/TPWRS.2022.3174922.
		\bibitem{16} \textcolor{black}{J. Wang, J. Zhang, Q. Hou, and N. Zhang, “Synchronous Condenser Placement for Multiple HVDC Power Systems Considering Short-Circuit Ratio Requirements,” {\it{IEEE Trans. Power Syst.}}, vol. 40, no. 1, pp. 765–779, Jan. 2025, doi: 10.1109/TPWRS.2024.3404116.}
	
		\bibitem{18} \textcolor{black}{China Power. (2021, Jul. 28). ``Action Plan for Building a New Type Power System with New Energy as the Main Body (2021–2030)." [Online]. Available:http://mm.chinapower.com.cn/xw/zyxw/20210728/90959.html.}
		\bibitem{22} \textcolor{black}{Polaris Transmission and Distribution Network. (2024, Oct. 11). “Ningxia’s First New Energy Distributed Synchronous Condenser Connected to Grid for Operation.” [Online]. Available: https://news.bjx.com.cn/html/20241011/1404359.shtml.}
		\bibitem{23} \textcolor{black}{Sohu. (2023,Dec.14). ``Synchronous Condenser: Effectively Supporting Power Grid and Promoting Green Power Consumption." [Online]. Available: https://www.sohu.com/a/743777154\_121124373.}
	
		\bibitem{25} \textcolor{black}{H. T. Nguyen, C. Guerriero, G. Yang, C. J. Boltonand, T. Rahman and P. H. Jensen, "Talega SynCon - Power Grid Support for Renewable-based Systems," in {\it{2019 SoutheastCon}}, Huntsville, AL, USA, 2019, pp. 1-6.}
		
		
		
		\bibitem{40} \textcolor{black}{M. Zheng, “The stability of synchronous condensers operating near a load center,” {\it{Proc. CSEE}}, no. 2, pp. 13–28, 1965.}
		\bibitem{26} B. Li, S. Yang, Y. Hu, Z. Hao, H. Xie, and T. Zhao, “Rotor Angle Transient First-Swing Stability Analysis of Synchronous Condensers Near Wind Farms,” in {\it{2023 IEEE PESGM}}, Orlando, FL, USA, Jul. 2023, pp. 1–5.
		\bibitem{27} \textcolor{black}{X. Liu, H. Xin, D. Zheng, D. Chen, and J. Tu, “Transient Stability of Synchronous Condenser Co-Located With Renewable Power Plants,” {\it{IEEE Trans. Power Syst.}}, vol. 39, no. 1, pp. 2030–2041, Jan. 2024, doi: 10.1109/TPWRS.2023.3271025.}
		\bibitem{28} X. Liu, H. Xin, Y. Shan, D. Zheng, and D. Chen, “Transient Stability of Synchronous Condenser Co-Located With Renewable Power Plants Under High-Resistance Faults and Risk Mitigation,” {\it{IEEE Trans. Sustain. Energy}}, vol. 15, no. 4, pp. 2581–2593, Oct. 2024, doi: 10.1109/TSTE.2024.3429210.
		\bibitem{12} \textcolor{black}{Y. Wang, H. Sun, S. Xu, and B. Zhao, “Transient Stability Analysis and Improvement for the Grid-Connected VSC System With Multi-Limiters,” {\it{IEEE Trans. Power Syst.}}, vol. 39, no. 1, pp. 1979–1995, Jan. 2024, doi: 10.1109/TPWRS.2023.3245806.}
		\bibitem{29} Y. Li, Y. Lu, and Z. Du, “Direct method of Lyapunov applied to synchronization stability of VSC with phase-locked loop,” {\it{Elect. Power Syst. Res.}}, vol. 220, Jul. 2023, Art. no. 109376, doi: 10.1016/j.epsr.2023.109376.
	
		\bibitem{31} \textcolor{black}{Z. Wang, L. Guo, X. Li, X. Zhou, J. Zhu and C. Wang, "Multi-Swing PLL Synchronization Transient Stability of Grid-Connected Paralleled Converters," {\it{IEEE Trans. Sustain. Energy}}, vol. 16, no. 1, pp. 716-729, Jan. 2025, doi: 10.1109/TSTE.2024.3481417.}
		
		
		
		\bibitem{32} \textcolor{black}{Y. Liu, H. Geng, C. He, W. Ding, C. Shen, and G. Yang, “Equivalent Aggregated Modeling of Multi-VSC System for Transient Synchronization Stability Analysis,” {\it{IEEE Trans. Power Syst.}}, vol. 39, no. 2, pp. 4296–4310, Mar. 2024, doi: 10.1109/TPWRS.2023.3311759.}
	
		\bibitem{36} \textcolor{black}{O. V. Gazizova, A. E. Morshchakin, and G. P. Kornilov, “Increasing the Static Stability of Synchronous Generators with Group Excitation Control,” in {\it{2024 UralCon}}, Ekaterinburg, Russia, Sep. 2024, pp. 817–821, doi: 10.1109/UralCon62137.2024.10718988.}
		\bibitem{37} S. Yang, B. Li, Z. Hao, Y. Hu, H. Xie, and T. Zhao, “Multi-Swing Transient Stability of Synchronous Generators and IBR Combined Generation Systems,” {\it{IEEE Trans. Power Syst.}}, vol. 40, no. 1, pp. 1144–1147, Jan. 2025, doi: 10.1109/TPWRS.2024.3460421.
		\bibitem{38} \textcolor{black}{C. Wu, Y. Lyu, Y. Wang, and F. Blaabjerg, “Transient Synchronization Stability Analysis of Grid-Following Converter Considering the Coupling Effect of Current Loop and Phase Locked Loop,” {\it{IEEE Trans. Energy Convers.}}, vol. 39, no. 1, pp. 544–554, Mar. 2024, doi: 10.1109/TEC.2023.3314095.}
		\bibitem{34} X. He and H. Geng, “Transient Stability of Power Systems Integrated With Inverter-Based Generation,” {\it{IEEE Trans. Power Syst.}}, vol. 36, no. 1, pp. 553–556, Jan. 2021, doi: 10.1109/TPWRS.2020.3033468.
		\bibitem{39} \textcolor{black}{P. Kokotovic, H. K. Khalil, and J. O’Reilly, {\it{Singular Perturbation Methods in Control: Analysis and Design}}. Philadelphia, PA, USA: SIAM, 1999.}
		\bibitem{41}
		\textcolor{black}{H.~K.~Khalil, \emph{Nonlinear Systems}, 3rd~ed. Upper Saddle River, NJ, USA: Prentice Hall, 2002.}

		
		\bibitem{42} \textcolor{black}{Y. Li, Y. Lu, Y. Tang and Z. Du, "Conditions of Existence and Uniqueness of Limit Cycle for Grid-Connected VSC With PLL," {\it{IEEE Trans. Power Syst.}}, vol. 39, no. 1, pp. 706-719, Jan. 2024, doi: 10.1109/TPWRS.2023.3238000.}		
\end{thebibliography}
\end{document}